\documentclass[journal]{IEEEtran}
\usepackage{xcolor,soul,framed} %,caption
\usepackage[pdftex]{graphicx}
\graphicspath{{../pdf/}{../jpeg/}}
\DeclareGraphicsExtensions{.pdf,.jpeg,.png}
\usepackage[cmex10]{amsmath}
\usepackage{amssymb} % use align

\usepackage{array}
\usepackage{mdwmath}
\usepackage{mdwtab}
\usepackage{eqparbox}
\usepackage{url}
\usepackage{subfigure}
\usepackage{float}
\RequirePackage{algorithm,algpseudocode}
\begin{document}
\bstctlcite{IEEEexample:BSTcontrol}
\title{Meta-Learning Empowered Graph Neural Networks for Radio Resource Management}
\author{
Kai~Huang,~\IEEEmembership{Graduate Student Member,~IEEE,}
Le~Liang,~\IEEEmembership{Member,~IEEE,}
Xinping~Yi,~\IEEEmembership{Member,~IEEE,}
Hao~Ye,~\IEEEmembership{Member,~IEEE,} Shi~Jin,~\IEEEmembership{Fellow,~IEEE}
and
Geoffrey~Ye~Li,~\IEEEmembership{Fellow,~IEEE}

\thanks{Kai Huang, Xinping Yi, and Shi Jin are with the National Mobile Communications Research Laboratory, Southeast University, Nanjing 210096, China (e-mail: hkk@seu.edu.cn; xyi@seu.edu.cn; jinshi@seu.edu.cn).}
\thanks{Le Liang is with the National Mobile Communications Research Laboratory, Southeast University, Nanjing 210096, China, and also with Purple Mountain Laboratories, Nanjing 211111, China (e-mail: lliang@seu.edu.cn).}
\thanks{Hao Ye is with the Department of Electrical and Computer Engineering, University of California, Santa Cruz, CA 95064, USA (e-mail: yehao@ucsc.edu).}
\thanks{Geoffrey Ye Li is with the ITP Lab, the Department of Electrical and Electronic Engineering, Imperial College London, SW7 2BX London, U.K. (e-mail: geoffrey.li@imperial.ac.uk).}
}  

% \markboth{Journal of \LaTeX\ Class Files,~Vol.~14, No.~8, July~2024}%

\maketitle

\begin{abstract}
In this paper, we consider a radio resource management (RRM) problem in the dynamic wireless networks, comprising multiple communication links that share the same spectrum resource. To achieve high network throughput while ensuring fairness across all links, we formulate a resilient power optimization problem with per-user minimum-rate constraints. We obtain the corresponding Lagrangian dual problem and parameterize all variables with neural networks, which can be trained in an unsupervised manner due to the provably acceptable duality gap. 
% Afterwards, we try to further improve the scalability of the proposed primal-dual method by adopting meta-learning. However, existing works suffer from the problem of time-varying dimensionality of NNs caused by the fast changes of the network configurations, e.g., the number of links, which makes the meta-learning framework infeasible. Therefore, graph neural networks (GNNs) are selected in this paper as parameterization due to their scalability, and a meta-learning empowered GNN algorithm is proposed. 
We develop a meta-learning approach with graph neural networks (GNNs) as parameterization that exhibits fast adaptation and scalability to varying network configurations. We formulate the objective of meta-learning by amalgamating the Lagrangian functions of different network configurations and utilize a first-order meta-learning algorithm, called Reptile, to obtain the meta-parameters. Numerical results verify that our method can efficiently improve the overall throughput and ensure the minimum rate performance. We further demonstrate that using the meta-parameters as initialization, our method can achieve fast adaptation to new wireless network configurations and reduce the number of required training data samples.
\end{abstract}

% KEYWORDS 
\begin{IEEEkeywords}
Radio resource management, adaptive power control, graph neural network, meta-learning, primal-dual problem 
\end{IEEEkeywords}

\IEEEpeerreviewmaketitle

% INTRODUCTION 

\section{Introduction}
% Background
\IEEEPARstart{W}{ith} the proliferation of the wireless devices and different quality-of-service (QoS) requirements, wireless networks are becoming larger and more complex. In order to improve the overall performance of the networks with limited wireless resources, novel radio resource management (RRM) methods should be developed. 

Power control constitutes an indispensable part of RRM problems. In this paper, we consider an adaptive power control problem in the dynamic wireless networks, which aims to maximize the sum ergodic rates of all communication links while ensuring fairness across all links. Such resource allocation problems are typically formulated as mathematical optimization problems, which are usually NP-hard and computationally challenging to derive the optimal solutions. 

Due to the success in computer vision and natural language processing, deep learning (DL) has been widely utilized in resource allocation problems, employing various network architectures such as multi-layer perceptron (MLP) \cite{sun2018learning} and convolutional neural networks (CNN) \cite{cui2019spatial}. Besides, deep reinforcement learning (DRL) has also exhibited its ability to deal with resource allocation problems \cite{liang2019spectrum}\cite{liang2020deep}. Recently, DL-based methods have been utilized in dealing with the considered adaptive power control problem. The formulated primal optimization problem with constraints in \cite{navid2022state} is turned into an unconstrained dual problem using the primal-dual method, and then a state-augmented method is proposed, in which the input of the neural networks (NNs) consists of instantaneous network state and the dual variables corresponding to the constraints. On the basis of \cite{navid2022state}, a resilient RRM problem with minimum-rate constraints is investigated in \cite{navid2023resilient}. By introducing slack terms, the constraints can be relaxed to ensure the feasibility of the optimization problem. Numerical results demonstrate that DL-based methods can achieve similar sum-rate performance with the baselines while outperforming them in terms of the minimum rate.

However, DL-based methods face challenges in the considered problem, where the wireless network configuration, including the number, topology, and fading states of the wireless links, changes frequently. Learning the NN parameters from scratch for one configuration usually requires cumbersome training with a large number of data samples while the obtained model may suffer performance degradation when confronting scenarios different from the one used for training. Moreover, it is difficult, if not impossible, in practice to obtain enough data samples for training in different network configurations. Motivated by the human ability to rapidly adapt to new tasks with experience in similar tasks, the generalization of DL-based resource management schemes can be improved via meta-learning \cite{finn2017model}. Gradient-based meta-learning is a critical part of meta-learning, which aims to obtain an initialization that enables the NNs to adapt to new tasks with minimal gradient descents and a limited number of data samples by training across many similar tasks \cite{chen2023learning}. Model-agnostic meta-learning (MAML) \cite{finn2017model} is a prominent meta-learning algorithm that has gained significant popularity due to its effectiveness in regression and deep reinforcement learning (DRL) tasks. Recently, meta-learning, particularly the MAML algorithm, has been employed in wireless communication to improve the generalization of the DL/DRL-based methods, such as beamforming \cite{wang2023new}, channel estimation \cite{wang2023learn},
%beam prediction \cite{yang2023meta}, trajectory design of unmanned aerial vehicles \cite{hu2021distributed} 
multiple-input multiple-output (MIMO) detection \cite{zhang2021meta}, spectrum sharing in vehicular networks \cite{huang2023meta}, etc. However, directly extending meta-learning to the considered problem presents challenges as meta-learning necessitates that all tasks be addressed by a dimension-invariant network model. Aforementioned MLP and CNN have poor scalability as their dimension relies heavily on the network size, in such a way that the network model should change once the network size changes. Existing works combining meta-learning and primal-dual problems, such as \cite{zhao2020primal} and \cite{zhao2020fair}, assume the dimensions of the dual variables are constant. However, when the network size changes, the dimensions of dual variables also change, rendering the methods in \cite{zhao2020primal} and \cite{zhao2020fair} infeasible. Currently, there are few effective approaches that combine meta-learning and primal-dual problems with dual variables of varying dimensions.

% Related Works
To solve this problem, graph neural networks (GNNs) are leveraged in this paper as the network models due to their desirable scalability and excellent performance. More recently, GNNs have been widely utilized in RRM problems. In \cite{shen2021graph}, the RRM problems are formulated as optimization over graphs, which enjoy the universal permutation equivariance property. As such, these graph optimization problems are solved by GNNs, which efficiently exploit the spatial structures and exhibit scalability. A link scheduling problem is considered in \cite{lee2021graph}, where the links are scheduled according to their graph embedding features to maximize the overall throughput while requiring fewer training samples. The more recent work for D2D link scheduling in \cite{shan2024GRLinQ}, GRLinQ, leverages message passing GNNs (MPGNNs) and reinforcement learning and achieves state-of-the-art throughput performance with excellent generalization and scalability. Besides, to relax the requirement of global channel state information (CSI), a sparse graph isomorphism network is proposed in \cite{wang2023sparse} by only reporting necessary CSI information via an edge-sparsifying mechanism. GNNs have been integrated with primal-dual method in \cite{liang2023dynamic} to effectively solve constrained optimization problems. GNN-based RRM methods have also been combined with meta-learning to enhance their generalization to different network configurations and various channel conditions \cite{zhao2024survey}. In \cite{ivana2023modular}, the integration of meta-learning and GNNs for power control maximizes the overall throughput of the wireless networks with an arbitrarily time-varying topology. For a similar problem formulation and wireless network scenario where the CSI distribution changes over time, the meta-gating framework in \cite{hou2023meta} considers the importance of each period and integrates them to achieve better performance. However, these meta-learning-based works only consider unconstrained optimization problems, and thus cannot address the dimensional changes of dual variables caused by changes in network size.

% Contribution
Therefore, we combine meta-learning with GNNs on the basis of prior works to effectively address constrained optimization problems in RRM and enhance generalization. Specifically, we propose a meta-learning empowered GNN method, called \textit{MetaGNN}. The main contributions of this paper can be summarized as follows:
\begin{itemize}
\item We first formulate the considered problem using the resilient RRM formulation in \cite{navid2023resilient} to make the optimization problem always feasible. Then this problem is solved by the primal-dual method. To address the dimensional changes caused by variations in the number of links, we utilize GNNs to parameterize the dual and slack variables of the resilient method. Under mild assumptions, we prove that this parameterization leads to an acceptable duality gap, hence enabling us to employ a gradient-based method to update the parameters.
\item We propose an algorithm based on meta-learning to obtain an efficient power control scheme rapidly with fewer samples in an unseen network configuration. The whole algorithm is divided into the meta-training and the adaptation stages, where the meta-training stage has a dual-loop structure. A GNN-based primal-dual learning method is used to obtain the parameters for specific network configurations in the inner loop. To decrease the computational complexity, we choose the first-order meta-learning method to update meta-parameters in the outer loop.
\item Based on simulation results, our algorithm is capable of acquiring a proficient initialization, with which the GNNs can adapt to new tasks with limited samples rapidly. We also show the superior performance of our scheme after adaptation compared to baseline methods. 
\end{itemize}

% Organization
The rest of the paper is organized as follows. In Sec. \ref{system model}, we introduce the system model and formulate the resilient optimization problem of power control in a dynamic wireless network. Then we transform the resilient problem to Lagrangian dual domain and propose a GNN-based primal-dual learning method in Sec. \ref{gnn algorithm}. Afterwards, the GNN-based primal-dual learning method is empowered by meta-learning in Sec. \ref{metagnn algorithm} to further improve the scalability. Simulation results are provided in Sec. \ref{numerical result} and the conclusion is drawn in Sec. \ref{conclusion}.
% System Model and Problem Formulation
\section{System Model and Problem Formulation}
\label{system model}
\begin{figure*}[ht!]
  \begin{center}  \includegraphics[width=0.88\textwidth]{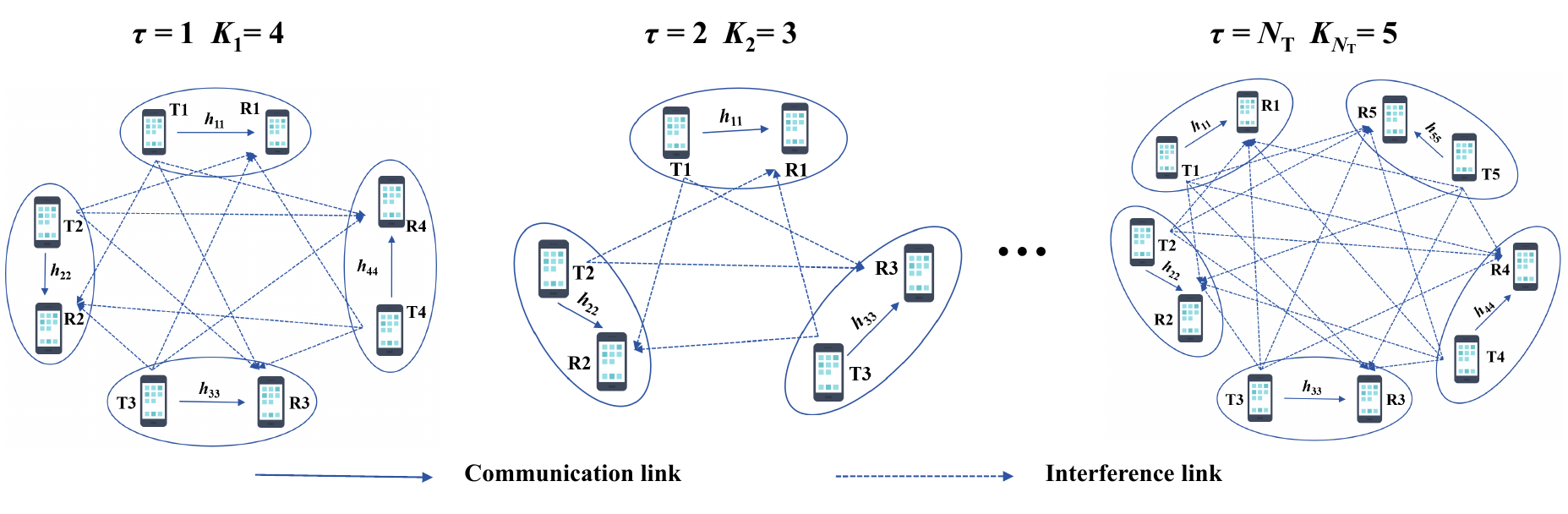}\\
  \caption{The dynamic wireless networks of different configurations $\tau$. The number of communication links, their locations and the fading states of the wireless links may change, leading to different network configurations.}
  \label{tasks}
  \end{center}
\end{figure*}
\begin{figure}[ht!]
  \begin{center}
  \includegraphics[width=0.48\textwidth]{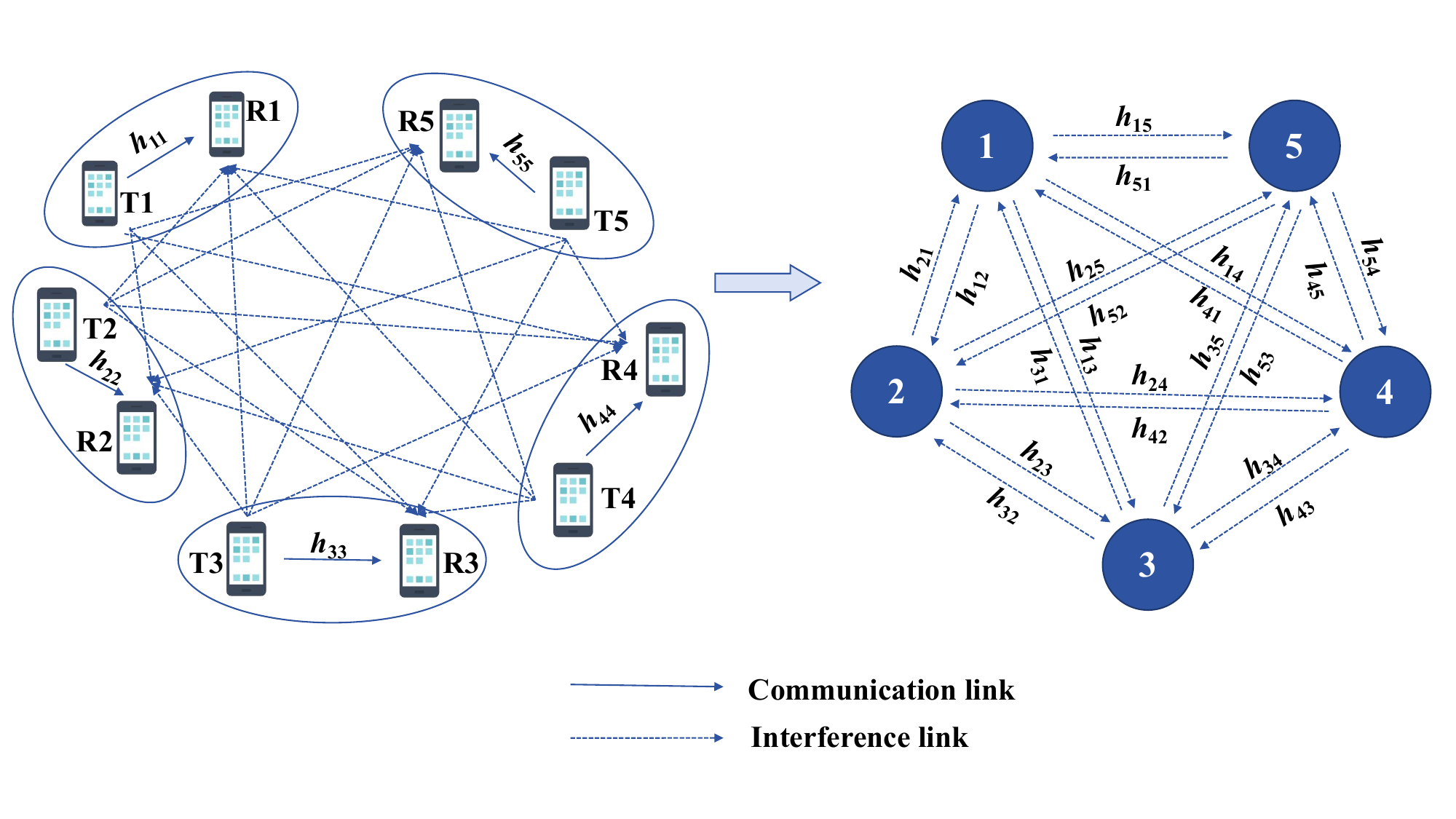}\\
  \caption{Graph modeling for the considered problem. Each node represents a communication link, and the directed edges mean the interfering links.}
  \label{graphmodel}
  \end{center}
\end{figure}

As illustrated in Fig. \ref{tasks}, we consider a dynamic wireless network with a number of paired transmitters and receivers, referred to as communication links, which join and leave the network in an autonomous manner. Therefore, the network size may change, forming a variety of different network configurations $\tau\in\lbrace1,\dots,N_T\rbrace$. For configuration $\tau$, there is a set of $K_{\tau}$ communication links denoted as $\mathcal{K}_{\tau}=\left\{1,\dots,K_{\tau}\right\}$. These links are assumed to transmit at the same time using the same spectrum, so they cause interference to each other. The channel gain of the $k$th link is denoted as $h_{kk}=\zeta_{kk}g_{kk}$, where $\zeta_{kk}$ is the frequency-independent large-scale fading component and $g_{kk}$ captures the small-scale fading effect. The interference channel gain from the transmitter of the $i$th link to the receiver of the $k$th link is represented as $h_{ik}$. All the channel gains across the wireless network are collected to form a matrix, ${\bf{H}}\in\mathbb{C}^{K_{\tau}\times K_{\tau}}$, which is drawn from the underlying distribution $\mathcal{U}_{\tau}({\bf{H}})$ related to the specific configuration of the network. 

The wireless network, as shown in Fig. \ref{tasks}, can be naturally modeled as an interference graph. Generally, a directed graph $\mathcal{G}(\mathcal{N}, \mathcal{E})$ consists of node set $\mathcal{N}$ and edge set $\mathcal{E}$. We denote the edge from node $i$ to node $j$ as $e_{i,j}\in\mathcal{E}$. In our setting, each communication link is considered as a node, and the interference link from the transmitter of the $i$th link to the receiver of the $j$th link is the edge $e_{i,j}$. In this way, the whole network can be modeled as a fully-connected graph model as shown in Fig. \ref{graphmodel}. For clarity of illustration, the directed edges are replaced by bidirected edges in the rest of the figures in this paper.

We consider an adaptive power control problem, where each communication link adjusts its transmit power in order to optimize a global network-wide objective. Given a fixed maximum transmit power $P_{\text{max}}$ and a channel gain matrix $\bf{H}\sim\mathcal{U}_{\tau}({\bf H})$, the transmit power vector of all links is denoted as ${\bf{p}_{\tau}(H)}\in[0, P_{\text{max}}]^{K_{\tau}}$. We can obtain the instantaneous signal-to-interference-plus-noise ratio (SINR) of the $k$th link in configuration $\tau$ as
\begin{align*} 
\text{SINR}_{k}({\bf {H}}, {\bf {p}}_{\tau}({\bf H})) = \frac{\left|h_{kk}\right|^{2} p_{k}}{\sigma^2 + \sum _{i=1, \;i\ne k}^{K_{\tau}} |h_{ik}|^{2} p_{i}}, 
\tag{1} 
\end{align*}
where $\sigma^2$ is the noise power and $p_i$, the $i$th component of $\bf{p}_{\tau}$, denotes the transmit power of the $i$th link. Then, we can obtain achievable rate of the $k$th link by
\begin{align*} 
f_{k}({\bf {H}}, {\bf {p}}_{\tau}({\bf H})) = \log_{2}(1 + \text{SINR}_{k}({\bf {H}}, {\bf {p}}_{\tau}({\bf H}))). 
\label{shannon}
\tag{2} 
\end{align*}

As the channel matrix varies over time, the transmit power vector should be adjusted accordingly. To capture the long-term overall throughput performance, we consider the ergodic rates of all links, which are denoted by ${\bf r}_{\tau}\in\mathbb{R}^{K_{\tau}}_{+}$. Thus, the power control problem with constrained ergodic rates can be stated as
\begin{align*} 
&\max _{{\bf {p}}_{\tau},{\bf {r}}_{\tau}} \hspace{15pt} {\bf {r}}_{\tau}^T\mathbf{1}_{K_{\tau}}, \tag{3a}\\ 
&\;\;\text{s.t.} \hspace{20pt} {\bf {r}}_{\tau}\leq \mathbb{E}_{{\bf H}\sim\mathcal{U}_{\tau}({\bf H})} \left[ {\bf {f}}({\bf {H}}, {\bf {p}}_{\tau}({\bf {H}})) \right],\tag{3b}\label{p1_b} \\ 
&\;\;\;\; \hspace{25pt} {\bf {r}}_{\tau}\geq {\bf {f}}_{\min }, \tag{3c}\label{p1_c}\\ 
&\;\;\;\; \hspace{25pt} {\bf {p}}_{\tau}({\bf {H}}) \in [0,P_{\max }]^{K_{\tau}},\tag{3d}\label{p1_d}
\end{align*}
where $\mathbf{1}_{K_{\tau}}$ is an all-ones vector with the dimension of $K_{\tau}$ and $\bf{H}$ is an instantaneous channel matrix. The objective is the ergodic sum rate, which is a concave operation. (\ref{p1_b}) implies that the ergodic rate of each link $r_{\tau,k}$ is limited by the ergodic Shannon capacity $\mathbb{E}_{{\bf {H}}\sim\mathcal{U}_{\tau}({\bf H})} \left[ f_k({\bf {H}}, {\bf {p}}_{\tau}({\bf {H}})) \right]$. In contrast to the sum-rate maximization problem in prior works \cite{ivana2023modular}\cite{hou2023meta}, we emphasize fairness across links by adding a constraint on the ergodic rate of the $k$th link to be no less than $f_{k, \text{min}}$. Besides, unlike prior works, such as \cite{liang2020towards}, that only consider the fairness across all links at each single time step, we consider the fairness in the long-term, which makes the problem more practical yet more challenging. This constraint mitigates power allocation bias in favour of better channel conditions, thereby promoting fairness across all links. Our goal is to adjust the transmit powers of all links so as to maximize the sum ergodic rate whilst meeting each link's minimum ergodic rate requirement simultaneously. 

However, solving the optimization problem in (3) requires prior knowledge of minimum rate constraints, i.e., ${\bf {f}}_{\min}$, for which the improper initializations may make the optimization problem infeasible for certain configurations. To resolve this issue, we consider the resilient RRM method in \cite{navid2023resilient} and introduce a slack term $\bf{s}_{\tau}$ to loose the constraint (\ref{p1_c}). At the same time, we add a penalty term, $\frac{\alpha}{2}\Vert {\bf {s}}_{\tau}\Vert _{2}^{2}$, to the objective function, where $\alpha>0$ is the weight to balance the sum ergodic rates and the minimum ergodic rate. The primal optimization problem then becomes
\begin{align*} 
P^* :=\hspace{5pt}&\max _{{\bf {p}}_{\tau},{\bf {r}}_{\tau},{\bf {s}}_{\tau}} \hspace{15pt} {\bf {r}}_{\tau}^T\mathbf{1}_{K_{\tau}} - \frac{\alpha}{2}\Vert {\bf {s}}_{\tau}\Vert _{2}^{2}, \tag{4a}\\ 
&\;\;\text{s.t.} \hspace{25pt} {\bf {r}}_{\tau}\leq \mathbb{E}_{{\bf H}\sim\mathcal{U}_{\tau}({\bf H})} \left[ {\bf {f}}({\bf {H}}, {\bf {p}}_{\tau}({\bf {H}})) \right],\tag{4b} \\ 
&\;\;\;\; \hspace{30pt} {\bf {r}}_{\tau}\geq {\bf {f}}_{\min } - {\bf {s}}_{\tau}, \tag{4c}\label{p2_c}\\ 
&\;\;\;\; \hspace{30pt} {\bf {p}}_{\tau}({\bf {H}}) \in [0,P_{\max }]^{K_{\tau}}, {\bf {s}}_{\tau}\geq {\bf 0}.\tag{4d}
\end{align*}
This problem is a relaxed version of (3) and is feasible because we can adjust $\bf{s}_{\tau}$ to ensure that (\ref{p2_c}) is always satisfied. It is worth noting that the optimal solutions of $\bf{r}_{\tau}$ and $\bf{s}_{\tau}$ are configuration-dependent. The current formulation is based on averaging over small-scale fading only. When the configurations change, the impact of large-scale fading, network size, topology, etc., need to be considered carefully. Therefore, our goals are mainly twofold. Firstly, we need to obtain the proper solution for problem (4) in configuration $\tau$, i.e., ${\bf{p}}_{\tau}$, ${\bf{r}}_{\tau}$, and ${\bf s}_{\tau}$. 
% This optimization problem involves both convex and non-convex constraints that cannot be directly solved. 
Secondly, we should design an effective adaptive power control scheme to optimize the performance of a family of network configurations, so that it can generalize across configurations once the training is finished. This scheme needs to provide universal models to deal with the environmental dynamics and utilize proper methods to improve the scalability. We will consider these two problems in detail in following sections, respectively. 

\section{GNN-based primal-dual learning method}\label{gnn algorithm}
In this section, we propose a GNN-based primal-dual learning method to obtain the proper solution of problem (4) in specific configuration $\tau$. In Sec. \ref{III-A}, we turn optimization problem (4) into Lagrangian dual domain and use the NNs to parameterize the mentioned optimization variables. Then in Sec. \ref{III-B}, GNNs are utilized to realize this parameterized primal-dual learning method thanks to the provable acceptable duality gap.
\subsection{Parameterized Primal-Dual Learning Method}
\label{III-A}
In order to deal with the constrained optimization problem in (4), we have to learn ${\bf{p}}_{\tau}$, ${\bf{r}}_{\tau}$, and ${\bf s}_{\tau}$ over a set of both convex and non-convex constraints. This can be achieved by transforming this problem into the Lagrangian dual domain. With two non-negative dual variables $\boldsymbol{\lambda}_{\tau},\boldsymbol{\mu}_{\tau}\in\mathbb{R}^{K_{\tau}}_{+}$, we can define the Lagrangian function as 
\begin{align*}
&{\mathcal {L}}_{\tau}({\bf {p}}_{\tau},{\bf {r}}_{\tau},{\bf {s}}_{\tau}, \boldsymbol{\lambda}_{\tau},\boldsymbol{\mu}_{\tau})= {\bf {r}}_{\tau}^T\mathbf{1}_{K_{\tau}} - \frac{\alpha}{2}\Vert {\bf {s}}_{\tau}\Vert _{2}^{2} \\ 
&\;\hspace{15pt} - \boldsymbol{\lambda }_{\tau}^{T}\left[{\bf {r}}_{\tau}- \mathbb{E}_{\bf{H}} \left[ {\bf {f}}({\bf {H}}, {\bf {p}}_{\tau}({\bf {H}})) \right] \right]- \boldsymbol{\mu }_{\tau}^{T} \left[ {\bf {f}}_{\min } - {\bf {s}}_{\tau}- {\bf {r}}_{\tau}\right], \tag{5}
\label{pri_lag}
\end{align*}
and obtain the dual problem
\begin{align*} 
D^* :=\hspace{5pt}\min _{\boldsymbol{\lambda }_{\tau}, \boldsymbol{\mu }_{\tau}} \max _{{\bf {p}}_{\tau}, {\bf {r}}_{\tau}, {\bf {s}}_{\tau}} {\mathcal {L}}_{\tau}({\bf {p}}_{\tau},{\bf {r}}_{\tau},{\bf {s}}_{\tau}, \boldsymbol{\lambda}_{\tau},\boldsymbol{\mu}_{\tau}), \tag{6} 
\label{dp}
\end{align*}
where we attempt to find the best primal variables ${\bf p}_{\tau}$, ${\bf r}_{\tau}$, ${\bf s}_{\tau}$, to maximize the Lagrangian function while seeking for the best dual variables $\boldsymbol{\lambda}_{\tau}$ and $\boldsymbol{\mu}_{\tau}$ to minimize it. With (\ref{pri_lag}), we can transform the optimization problem with constraints into a simple unconstrained problem. It is worthy of noting that the optimal solution to dual problem (\ref{dp}), $D^*$, is the same as the optimal solution of primal problem (4), $P^*$, because of the zero-duality gap \cite{eisen2019optimal}.

In fact, ${\bf p}_{\tau}$, ${\bf r}_{\tau}$, ${\bf s}_{\tau}$, $\boldsymbol{\lambda}_{\tau}$ and $\boldsymbol{\mu}_{\tau}$ can be considered as mappings from the environmental information, such as channel matrix $\bf H$, to corresponding outputs. Consequently, (\ref{dp}) is a functional optimization problem aiming to seek the optimal mapping within an infinite-dimensional function space. In order to avoid tackling the infinite-dimensional optimization problem, we parameterize these mapping functions, thereby restricting the original function space to the parameter space and transforming (\ref{dp}) into a finite-dimensional optimization problem. Given that the previous works \cite{hornik1989multilayer}-\cite{cybenko1989approximation} have mentioned that DNNs can approximate arbitrary functions, we use them to represent the primal and dual functions. The power control function ${\bf p}_{\tau}$, the ergodic rate function ${\bf r}_{\tau}$, the slack function ${\bf s}_{\tau}$, and the dual functions $\boldsymbol{\lambda}_{\tau}$, $\boldsymbol{\mu}_{\tau}$ are replaced with finite-dimensional parameter vectors ${\bf p}_{\tau}(\cdot;\boldsymbol{\theta}^{\bf p})$, ${\bf{r}}_{\tau}(\cdot; \boldsymbol{\theta}^{{\bf{r}}})$, ${\bf{s}}_{\tau}(\cdot; \boldsymbol{\theta}^{{\bf{s}}})$, $\boldsymbol{\lambda}_{\tau}(\cdot; \boldsymbol{\theta}^{\boldsymbol{\lambda}})$, and $\boldsymbol{\mu}_{\tau}(\cdot; \boldsymbol{\theta}^{\boldsymbol{\mu}})$, and they are named as the power net, ergodic rate net, slack net and dual nets, respectively. Different from \cite{navid2022state} and \cite{navid2023resilient}, which only parameterize the power control function ${\bf p}_{\tau}$ and may hinder the scalability, we parameterize the dual and slack functions as well to facilitate transferring our method to various networks. The parameterized Lagrangian function then becomes
\begin{align*} 
&\mathcal{L}_\theta \left(\boldsymbol{\theta}^{\bf{p}},\boldsymbol{\theta}^{\bf {r}},\boldsymbol{\theta}^{\bf {s}}, \boldsymbol{\theta}^{\boldsymbol{\lambda }},\boldsymbol{\theta}^{\boldsymbol{\mu}}\right) \\
&\;= {\mathcal {L}}_{\tau} \left({\bf {p}}_{\tau}(\cdot; \boldsymbol{\theta }^{{\bf {p}}}), {\bf{r}}_{\tau}(\cdot; \boldsymbol{\theta}^{{\bf{r}}}), {\bf{s}}_{\tau}(\cdot; \boldsymbol{\theta}^{{\bf{s}}}), \boldsymbol{\lambda}_{\tau}(\cdot; \boldsymbol{\theta}^{\boldsymbol{\lambda}}),\boldsymbol{\mu}_{\tau}(\cdot; \boldsymbol{\theta}^{\boldsymbol{\mu}})\right), \tag{7} 
\label{p-lag}
\end{align*}
and the parameterized dual problem (\ref{dp}) changes to
\begin{align*} 
D_{\boldsymbol{\theta}}^* :=\min _{\boldsymbol{\theta}^{\boldsymbol{\lambda }}, \boldsymbol{\theta}^{\boldsymbol{\mu}}} \max _{\boldsymbol{\theta}^{{\bf {p}}}, \boldsymbol{\theta}^{\bf {r}}, \boldsymbol{\theta}^{\bf {s}}} {\mathcal {L}}_\theta \left(\boldsymbol{\theta}^{\bf{p}},\boldsymbol{\theta}^{\bf {r}},\boldsymbol{\theta}^{\bf {s}}, \boldsymbol{\theta}^{\boldsymbol{\lambda }},\boldsymbol{\theta}^{\boldsymbol{\mu}}\right). \tag{8} 
\label{dp_para}
\end{align*}
Here, the input of the parameterized power function ${\bf {p}}_{\tau}(\cdot; \boldsymbol{\theta}^{{\bf {p}}})$ is the instantaneous channel matrix $\bf H$, while the input of the other parameterized functions is the configuration information denoted as ${\bf{N}}$, such as the realization of the placement of the communication links, the corresponding long-term fading state, etc.

Since we have parameterized dual problem (\ref{dp}), problem (\ref{dp_para}) no longer holds zero-duality gap property, i.e., $D^*_{\theta}\neq P^*$. In what follows, we show that the duality gap between the primal and the parameterized dual problems is upper bounded under mild conditions.
\begin{figure*}[ht!]
  \begin{center}
  \includegraphics[width=0.75\textwidth]{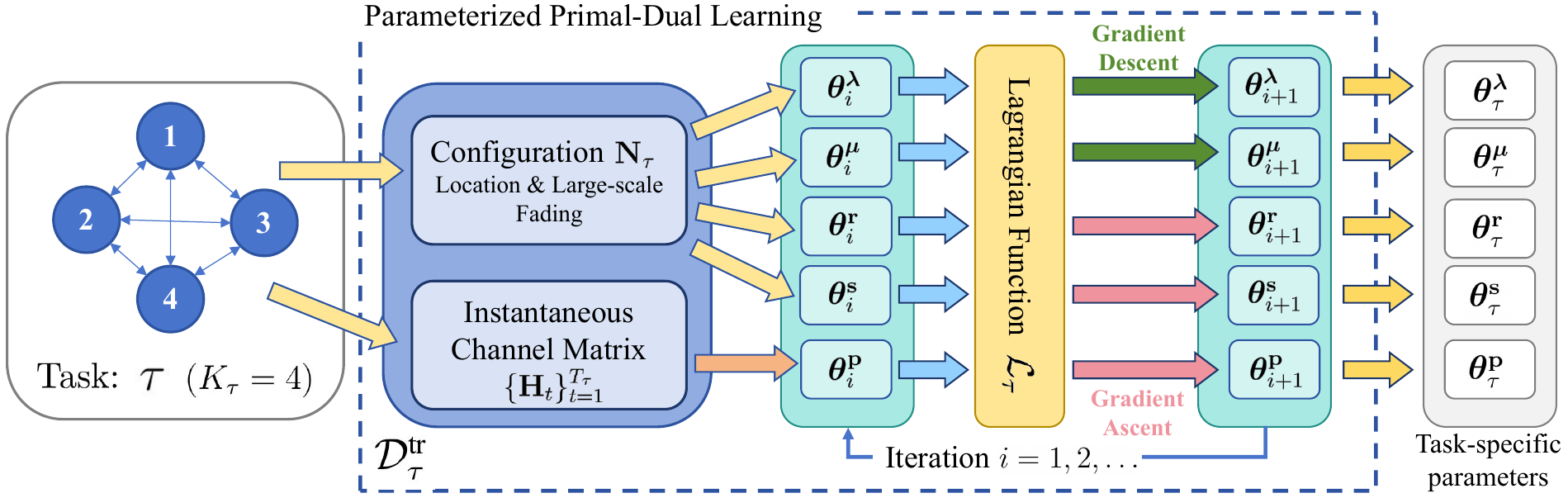}\\
  \caption{Framework of the parameterized primal-dual learning method for a specific task $\tau$. The bidirected edges mean two communication links are interfering each other.}
  \label{parameterizedduallearning}
  \end{center}
\end{figure*}

\textit{\textbf{Assumption 1:}} The probability distribution, $\mathcal{U}_{\tau}({\bf H})$, is non-atomic in $\mathcal{H}$, i.e., for any set $\mathcal{X}\subseteq \mathcal{H}$ of non-zero probability, there exists a non-zero probability strict subset $\mathcal{X}^{'}\subset\mathcal{X}$ of lower probability, $0<\mathbb{E}_{\bf H}(\mathbb{I}(\mathcal{X}^{'}))<\mathbb{E}_{\bf H}(\mathbb{I}(\mathcal{X}))$, where $\mathbb{I}(\cdot)$ denotes the indicator function.

\textit{\textbf{Assumption 2:}} For problem (4), Slater's condition holds. That is, there exist variables ${\bf r}_{\tau}^0$, ${\bf p}_{\tau}^0({\bf H})$ and a strictly positive scalar constant $\sigma>0$ such that
\begin{equation*}
    \mathbb{E}_{{\bf H}}\left[{\bf f}({\bf H}, {\bf p}_{\tau}^0({\bf H}))\right]-{\bf r}_{\tau}^0\geq\sigma\mathbf{1}.
    \tag{9}
\end{equation*}

\textit{\textbf{Assumption 3:}} The expected performance function, $\mathbb{E}_{{\bf H}}\left[{\bf f}({\bf H}, {\bf p}_{\tau}({\bf H}))\right]$, is expectation-wise Lipschitz on ${\bf p}_{\tau}({\bf H})$ for all channel realizations $\bf H$, i.e., $\forall {\bf p}_{\tau}^1({\bf H}), {\bf p}_{\tau}^2({\bf H}) \in [0, P_{\text{max}}]^{K_{\tau}}$, there is a constant $L_f$ such that 
\begin{align*}
&\mathbb{E}_{\bf H}\Vert {\bf {f}}({\bf {H}}, {\bf {p}}_{\tau}^{1}({\bf {H}})) - {\bf {f}}({\bf {H}}, {\bf {p}}_{\tau}^2({\bf {H}})) \Vert _{\infty }\\ &\quad\quad\quad\quad\quad\quad \leq L_{f} \mathbb{E}_{\bf H}\Vert {\bf {p}}_{\tau}^{1}({\bf {H}}) - {\bf {p}}_{\tau}^{2}({\bf {H}})\Vert _{\infty }. \tag{10}
\end{align*}

\textit{\textbf{Assumption 4:}} The parameterization functions ${\bf p}_{\tau}(\cdot;\boldsymbol{\theta}^{\bf p})$ satisfies $\epsilon_p$-universal parameterization theorem in \cite{eisen2019optimal}, i.e., for any possible power function ${\bf p}_{\tau}$, we can always find $\boldsymbol{\theta}^{\bf p} \in\Theta$ such that
\begin{equation*}
    \mathbb{E}_{{\bf H}}\Vert {\bf p}_{\tau}({\bf H}) - {\bf p}_{\tau}({\bf H};\boldsymbol{\theta}^{\bf p})\Vert_{\infty}\leq\epsilon_p.
    \tag{11}
\end{equation*}
This assumption is weaker than the universal approximation theorem mentioned below. Specifically, if a parameterization function satisfies the universal-approximation theorem, then it also satisfies Assumption 4.

\textit{\textbf{Assumption 5:}}
The optimal parameterized functions ${\bf r}_{\tau}({\bf N};\boldsymbol{\theta}^{\bf r})$ and ${\bf s}_{\tau}({\bf N};\boldsymbol{\theta}^{\bf s})$ satisfy the universal-approximation theorem proposed in \cite{hornik1989multilayer}-\cite{cybenko1989approximation}, i.e.,  $\forall{{\bf r}_{\tau}, {\bf s}_{\tau}}$, $\exists \boldsymbol{\theta}^{\bf r},\boldsymbol{\theta}^{\bf s}\in\Theta$ that  for any ${\bf N}$, there are constants $\epsilon_r$ and $\epsilon_s$ such that
\begin{align*}
&\Vert {\bf r}_{\tau}({\bf N}) - {\bf r}_{\tau}({\bf N};\boldsymbol{\theta}^{\bf r})\Vert_{\infty}\leq\epsilon_r, \tag{12}\\
&\Vert {\bf s}_{\tau}({\bf N}) - {\bf s}_{\tau}({\bf N};\boldsymbol{\theta}^{\bf s})\Vert_{\infty}\leq\epsilon_s. \tag{13}
\end{align*}

Assumption 1 guarantees that in  $\mathcal{U}_{\tau}({\bf H})$, there is no point with strictly positive probability, given that the channel matrices observed in practice are drawn from a continuous distribution. Assumption 2 demonstrates that service demands can be provisioned with some slack, which is realizable in (4) via sufficiently small $\bf r$ and large $\bf s$. Assumption 3 is a continuity statement on each of the dimensions of the expectations of $\bf f$, which is weaker than general Lipschitz continuity. Assumptions 4-5 ensure that the optimal solutions to the parameterized dual problem in (\ref{dp_para}) approximate the corresponding dual problem in (4) well. According to the universal approximation theorem \cite{hornik1989multilayer}, there exist neural networks that meet Assumptions 4 and 5. Under Assumptions 1-5, we demonstrate that the duality gap $|D_{\theta}^*-P^*|$ is within an acceptable range, as stated below, and proved in Appendix \ref{proof}. 

\textit{\textbf{Theorem 1:}} Consider the primal power control problem in (4) and its parameterized dual problem (\ref{dp_para}). If Assumptions 1-5 hold, then optimal dual value $D_{\theta}^*$ satisfies
\begin{align*}
0\le P^*-D_{\theta}^* \le D(K, \boldsymbol{\gamma}^*),
\label{duality gap}
\tag{14}
\end{align*}
where $D(K, \boldsymbol{\gamma}^*) = \epsilon _{p} L_{f}\left\Vert \boldsymbol{\gamma}^*\right\Vert _{1}+(K+\Vert\boldsymbol{\gamma}^*\Vert_{1})\epsilon_r+ (\alpha Kf_{\min}+\Vert\boldsymbol{\gamma}^*\Vert_{1})\epsilon_s$ and $\boldsymbol{\gamma}^*=\left[\boldsymbol{\lambda}^*,\boldsymbol{\mu}^*\right]$ denotes the optimal value of these two dual variables. 

This theorem allows us to directly solve the finite-dimensional unconstrained problem (\ref{dp_para}) to obtain a near-optimal solution of the primal optimization problem in (4). However, the Lagrangian function shown in (\ref{p-lag}) cannot be exactly computed in practice as the underlying distribution of the channel matrix is unknown. Therefore, we resort to an empirical Lagrangian function, replacing the expectation over $\mathcal{U}_{\tau}({\bf H})$ by an empirical average value, which is written as 
\begin{align*} 
&\hat{{\mathcal{L}}}_\theta \left(\boldsymbol{\theta}^{\bf{p}},\boldsymbol{\theta}^{\bf {r}},\boldsymbol{\theta}^{\bf {s}}, \boldsymbol{\theta}^{\boldsymbol{\lambda }},\boldsymbol{\theta}^{\boldsymbol{\mu}}\right) = {\bf {r}}_{\tau}({\bf{N}}; \boldsymbol{\theta}^{\bf {r}})^T\mathbf{1}_{K_{\tau}} - \frac{\alpha }{2}\Vert {\bf {s}}_{\tau}({\bf{N}}; \boldsymbol{\theta}^{\bf {s}})\Vert _{2}^{2} \\ 
&\;\quad \, - \boldsymbol{\lambda}_{\tau}({\bf{N}};\boldsymbol{\theta}^{\boldsymbol{\lambda}})^{T}\left[{\bf {r}}_{\tau}({\bf{N}}; \boldsymbol{\theta}^{\bf {r}})- \hat{\mathbb{E}}_{{\bf {H}}} \left[ {\bf {f}}({\bf {H}}, {\bf {p}}_{\tau}({\bf {H}};\boldsymbol{\theta }^{{\bf {p}}})) \right] \right] \\ 
&\;\quad\, -\boldsymbol{\mu}_{\tau}({\bf{N}};\boldsymbol{\theta}^{\boldsymbol{\mu}})^{T} \left[ {\bf {f}}_{\min } - {\bf {s}}_{\tau}({\bf{N}}; \boldsymbol{\theta}^{\bf {s}})- {\bf {r}}_{\tau}({\bf{N}}; \boldsymbol{\theta}^{\bf {r}})\right], \tag{15}\label{ep_lag} 
\end{align*}
where
\begin{align*} 
\hat{\mathbb{E}}_{{\bf {H}}} \left[{\bf {f}}({\bf {H}}, {\bf {p}}_{\tau}({\bf {H}};\boldsymbol{\theta }^{{\bf {p}}}))\right] &:=\frac{1}{T_{\tau}} \sum _{t=1}^{T_{\tau}} {\bf {f}}({\bf {H}}_t, {\bf {p}}_{\tau}({\bf {H}}_t;\boldsymbol{\theta }^{{\bf {p}}})). \tag{16} 
\end{align*}
The channel matrix samples, $\lbrace{\bf H}_t\rbrace_{t=1}^{T_{\tau}}$, are drawn from the underlying distribution. After obtaining the empirical Lagrangian function (\ref{ep_lag}), we can adopt a stochastic gradient update method. Iteration index $i$ indicates the number of times that parameters have updated. For the power net, its parameters are updated according to
\begin{align*} 
\boldsymbol{\theta}^{{\bf {p}}}_{i+1} = \boldsymbol{\theta }^{{\bf {p}}}_{i} + \eta _{p} \nabla _{\boldsymbol{\theta }^{{\bf {p}}}_{i}} \lbrace \boldsymbol{\lambda}_{\tau}({\bf N})^{T} \hat{\mathbb{E}}_{{\bf {H}}} \left[ {\bf {f}}({\bf {H}}, {\bf {p}}_{\tau}({\bf {H}};\boldsymbol{\theta }^{{\bf {p}}}_{i})) \right] \rbrace,  \tag{17}\label{updatep}
\end{align*}
where $\eta_p>0$ denotes the learning rate. The parameters of the ergodic rate net and slack net are updated using gradient ascent
\begin{align*} 
\boldsymbol{\theta}^{{\bf r}}_{i+1} = \boldsymbol{\theta}^{{\bf r}}_{i} + \eta_{r}\nabla _{\boldsymbol{\theta}^{{\bf r}}_{i}} \lbrace {\bf {r}}_{\tau}({\bf{N}}; \boldsymbol{\theta}^{{\bf r}}_{i})^T(\mathbf{1} + \boldsymbol{\mu}_{\tau}({\bf N}) - \boldsymbol{\lambda}_{\tau}({\bf N}))\rbrace, \tag{18} \label{updater}
\end{align*}
and
\begin{align*}
\boldsymbol{\theta }^{\bf {s}}_{i+1} = \boldsymbol{\theta }^{\bf {s}}_{i} + \eta_{s}\nabla _{\boldsymbol{\theta }^{\bf {s}}_{i}}\lbrace \boldsymbol{\mu}_{\tau}({\bf N})-\alpha{\bf {s}}_{\tau}({\bf{N}}; \boldsymbol{\theta }^{\bf {s}}_{i})\rbrace, \tag{19}\label{updates}
\end{align*}
where $\eta_r$ and $\eta_s$ are the learning rates of the ergodic rate and slack networks, respectively. These three networks are called primal networks. For the parameters of the dual networks, we adopt gradient descent method as
\begin{align*} 
\boldsymbol{\theta }^{\boldsymbol{\lambda}}_{i+1} = \boldsymbol{\theta }^{\boldsymbol{\lambda}}_{i} - \eta_{\lambda }\nabla _{\boldsymbol{\theta }^{\boldsymbol{\lambda}}_{i}}\Big\lbrace - &\boldsymbol{\lambda}_{\tau}({\bf{N}};\boldsymbol{\theta }^{\boldsymbol{\lambda}}_{i})^{T}\\
&\cdot\left[{\bf {r}}_{\tau}({\bf N})- \hat{\mathbb{E}}_{{\bf {H}}} \left[ {\bf {f}}({\bf {H}}, {\bf {p}}_{\tau}({\bf {H}}) \right] \right] \Big\rbrace, \tag{20} \label{updatel}
\end{align*}
and
\begin{align*}
\boldsymbol{\theta }^{\boldsymbol{\mu}}_{i+1} = \boldsymbol{\theta }^{\boldsymbol{\mu}}_{i}-\eta_{\mu}\nabla _{\boldsymbol{\theta }^{\boldsymbol{\mu}}_{i}}\Big\lbrace-&\boldsymbol{\mu}_{\tau}({\bf{N}};\boldsymbol{\theta }^{\boldsymbol{\mu}}_{i})^T\\
&\cdot\left[ {\bf {f}}_{\min } - {\bf {s}}_{\tau}({\bf N})- {\bf {r}}_{\tau}({\bf N})\right]\Big\rbrace, \tag{21}\label{updatem}
\end{align*}
where $\eta_{\lambda}$ and $\eta_{\mu}$ are the learning rates of these two dual networks. When the parameters finally converge, we extract them as task-specific parameters, denoted as $\boldsymbol{\theta}^{\bf I}_{\tau}, {\bf I} \in\{{\bf p}, {\bf r}, {\bf s},\boldsymbol{\lambda}, \boldsymbol{\mu}\}$. The whole process is shown in Fig. \ref{parameterizedduallearning}. 

\subsection{The Realization of Parameterization: GNNs}
\label{III-B}
The selection of the parameterization significantly affects the performance of the final solution to (\ref{dp_para}). Since we have modeled the wireless networks as graphs, utilizing GNNs as parameterization is a natural choice. Comparing to DNNs, GNNs offer several advantages in addressing the power control problem within wireless networks. Firstly, GNNs have the ability to effectively capture the topological information of the interference network. Secondly, GNNs have permutation equivariance that means the variation of link indices does not affect the power control schemes. Lastly, network parameters are shared across nodes in GNNs, allowing for the direct deployment upon the addition of new nodes, which shows better scalability than DNNs that require re-training. GNNs are known to satisfy the near-universality in Assumption 4 for the class of continuous, equivariant functions \cite{keriven2019universal}. Moreover, there are a considerable amount of researches investigating the expressive power of GNNs, such as \cite{xu2018powerful} and \cite{azizian2021expressive}, demonstrating that GNNs exhibit properties similar to the universal approximation theorem. Therefore, we also consider GNNs to satisfy Assumptions 5 and we demonstrate that using GNNs as parameterization incurs only limited performance loss. 

In this paper, typical MPGNNs are used as the backbone for parameterization. The update of the $m$th node in the $l$th layer of the MPGNN is denoted as
\begin{equation*}
{\bf {x}}_{m}^{l}=\text{MLP}_1^l\left({\bf {x}}_{m}^{l-1},\varphi_{j\in\mathcal{N}_m}\lbrace\text{MLP}_2^l[{\bf x}_{j}^{l-1}, e_{jm}]\rbrace\right),\tag{22}
\end{equation*}
where ${\bf {x}}_{m}^{l}\in\mathbb{R}^{D_l}$ denotes the node embedding of the $m$th node in the $l$th layer with $D_l$ representing its dimension. ${\bf {x}}_{m}^{0}$ is the initial node feature and $\mathcal{N}_m$ is the set of neighbors of node $m$. $\text{MLP}_2^l(\cdot)$ performs message computation and $\varphi(\cdot)$ is commonly implemented by maximum or summation function. $\text{MLP}_1^l(\cdot)$ combines the aggregated information from neighbors and its own information. After $L$ aggregation layers, the final node embeddings for all the nodes in the graph is ${\bf {x}}^{(L)}\in\mathbb{R}^{D_L\times K}$. The node embeddings can be further processed with extra MLPs to obtain the corresponding transmit power, ergodic rate, slack and dual values, which is described in details below.
\begin{figure}[ht!]
  \begin{center}
  \includegraphics[width=0.49\textwidth]{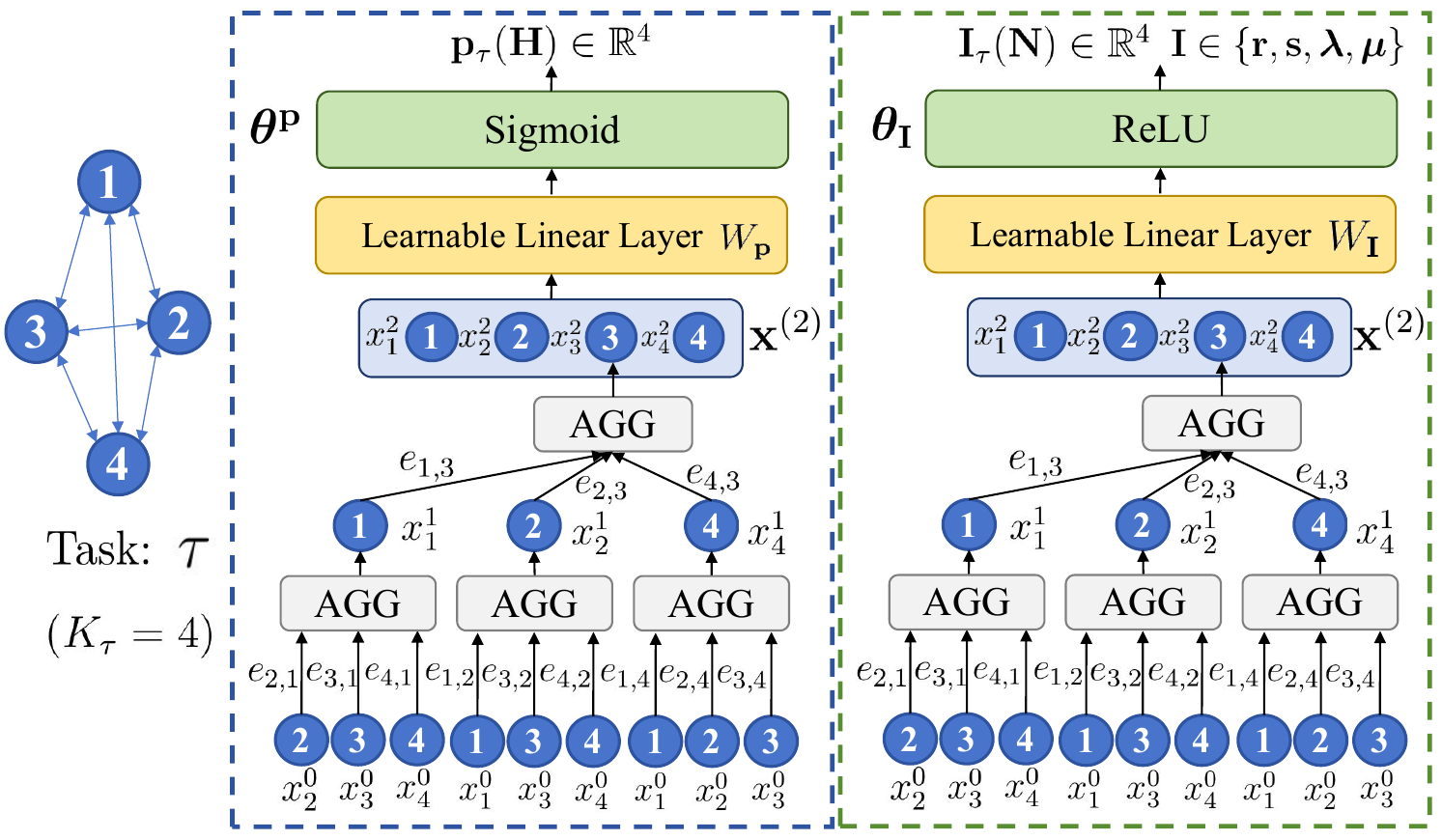}\\
  \caption{The GNN framework of these parameterized functions, where $L=2$. The left one is the power net, of which the input is the instantaneous channel matrix $\bf H$. The right one shows the framework of ergodic rate, slack and dual nets, of which the input is the configuration information ${\bf N}$, i.e., large-scale fading information.}
  \label{graphframework}
  \end{center}
\end{figure}
\subsubsection{Power net}
For all communication links, the transmission power is generated as
\begin{equation*}
    {\bf p}_{\tau}({\bf H})=P_{\max}\cdot\text{Sigmoid}(W_{\bf p}^T{\bf {x}}^{(L)}),\tag{23}
\end{equation*}
where $W_{\bf p}\in\mathbb{R}^{D_L}$ is a learnable linear layer. Due to the sigmoid function, the generated power values satisfy the constraint innately. As shown in Fig. \ref{parameterizedduallearning}, the input of the power net is the instantaneous channel matrix, ${\bf H}_t$, that consists of large-scale and small-scale fading components. The edge weights are set as the normalized channel gains in dB. For example, the weight of the edge from node $k$ to node $j$ is  
\begin{equation*}
    \mathcal{W}(e_{k,j})=\frac{\log(P_{\max}|h_{kj}|^2/\sigma^2)}{\sqrt{\sum_{k^{'},j^{'}\in\mathcal{N}}[\log(P_{\max}|h_{k^{'}j^{'}}|^2/\sigma^2)]^2}}. \tag{24}
\end{equation*}
We adopt the proportional-fairness (PF) ratio utilized in \cite{navid2023resilient} as the initial node features, i.e., ${\bf {x}}_{k}^{0}(t)=\text{PF}_k(t)$. The PF ratio $\text{PF}_k(t)$ is defined as 
\begin{equation*}
    \text{PF}_k(t) = \hat{r}_k(t)/\tilde{r}_k(t),
    \tag{25}
\end{equation*}
 where $\hat{r}_k(t)=f_k({\bf H}_t, P_{\max}1_{K_{\tau}})$ is the ratio of the estimated rate of the $k$th link at current time step $t$, $\tilde{r}_k(t)=(1-\beta)\tilde{r}_k(t-1)+\beta f_k({\bf H}_t, {\bf p}_{\tau}({\bf H}_t))$ denotes the exponential moving-average rate for $t\in\{1,2,\dots\}$, and $\beta\in[0, 1]$ is the averaging window length. $\tilde{\bf r}(0)$ denotes the initial long-term average rates which is pre-calculated using several data samples.
 
\begin{figure*}[ht!]
  \begin{center}
  \includegraphics[width=0.88\textwidth]{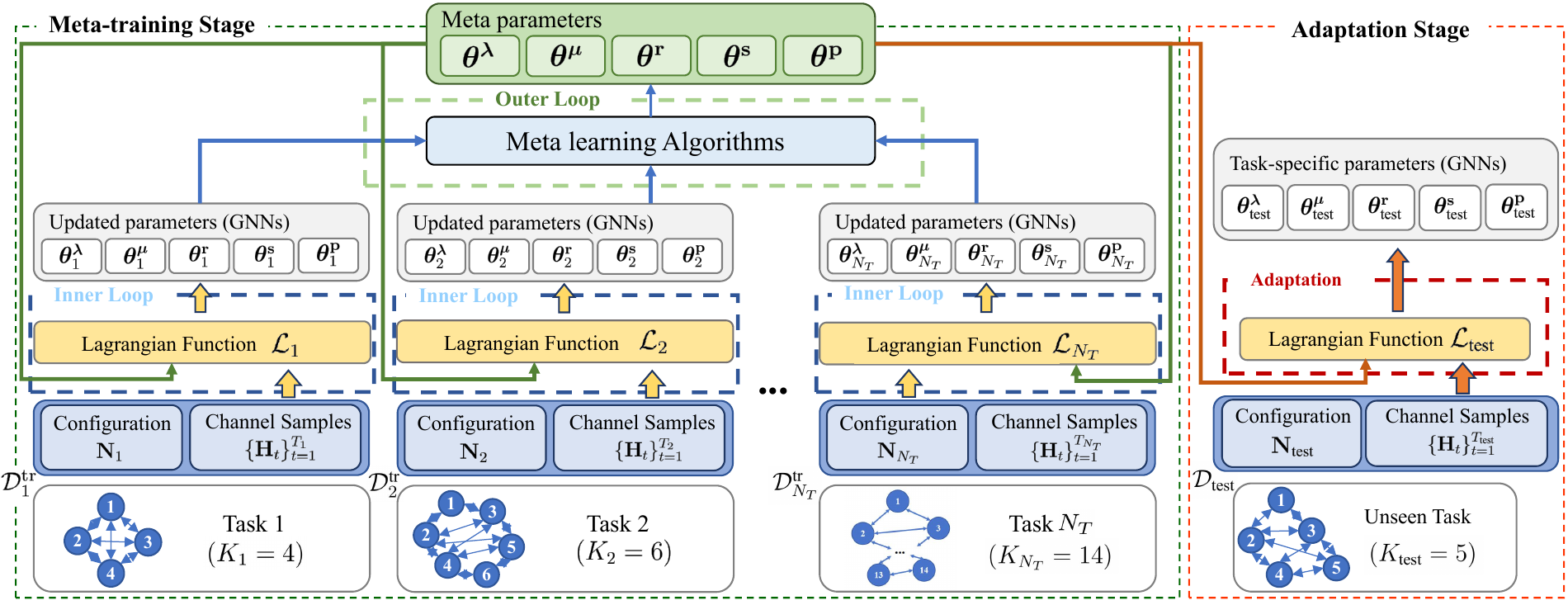}\\
  \caption{Structure of the meta-training and adaptation stages of meta-learning empowered GNN approach.}
  \label{metapro}
  \end{center}
\end{figure*}
\begin{figure*}[htb] % hb底部，ht为头部
\centering % 公式居中
\begin{align*} 
\label{metaobjective}
\min _{\boldsymbol{\theta}^{\boldsymbol{\lambda }}, \boldsymbol{\theta}^{\boldsymbol{\mu}}} \max _{\boldsymbol{\theta}^{{\bf {p}}}, \boldsymbol{\theta}^{\bf {r}}, \boldsymbol{\theta}^{\bf {s}}} \sum_{\tau=1}^{N_T} \mathcal{L}_{\tau}(&\boldsymbol{\theta}^{{\bf {p}}}+\eta_p \nabla _{\boldsymbol{\theta }^{{\bf {p}}}}\mathcal{L}_{\tau}(\boldsymbol{\theta}),\boldsymbol{\theta}^{\bf {r}}+\eta_r\nabla_{\boldsymbol{\theta}^{\bf r}}\mathcal{L}_{\tau}(\boldsymbol{\theta}), \boldsymbol{\theta}^{\bf {s}}+\eta_s\nabla_{\boldsymbol{\theta}^{\bf s}}\mathcal{L}_{\tau}(\boldsymbol{\theta}), \boldsymbol{\theta}^{\boldsymbol{\lambda }}-\eta_{\lambda}\nabla_{\boldsymbol{\theta}^{\boldsymbol{\lambda }}}\mathcal{L}_{\tau}(\boldsymbol{\theta}), \boldsymbol{\theta}^{\boldsymbol{\mu }}-\eta_{\mu}\nabla_{\boldsymbol{\theta}^{\boldsymbol{\mu }}}\mathcal{L}_{\tau}(\boldsymbol{\theta})). \tag{28}
\end{align*}
\hrulefill % 添加一条水平线
\vspace*{3pt} % 调整线与公式之间的距离
\end{figure*}

\subsubsection{Dual net, ergodic rate net and slack net}
As shown in Fig. \ref{parameterizedduallearning}, the input of these networks is the configuration of current wireless network, $\bf N$. The configuration primarily refers to the locations of the transmitters and the receivers. To facilitate the use of GNNs, we replace the locations with large-scale fading information as it implicitly contains the location information. The outputs of these networks will be processed via the ReLU activation function to ensure the non-negative property, i.e.,
\begin{equation*}
    {\bf I}_{\tau}({\bf N})=\text{ReLU}(W_{\bf I}^T{\bf {x}}^{(L)}), \quad{\bf I}\in\lbrace{\bf r, s}, \boldsymbol{\lambda, \mu}\rbrace, \tag{26}
\end{equation*}
where $W_{\bf I}\in\mathbb{R}^{D_L}$ is a learnable linear layer for each network. The edge weights are set as the normalized large-scale fading gains in dB. The initial node features can be set as pre-calculated exponential moving-average rate $\tilde{\bf r}(0)$ within several provided channel samples. In fact, the initial features only need to be fixed values as all information about the configuration is already encapsulated in the edge weights and these networks can extract useful features and obtain proper outputs. The frameworks of these nets are shown in Fig. \ref{graphframework}.

Moreover, the weight initialization of $W_{\bf I}$, ${\bf I}\in\lbrace{\bf r, s}, \boldsymbol{\lambda, \mu}\rbrace$, is crucial to the output of these networks. If the weight of $W_{\bf I}$ is conventionally zero-mean initialized, the network may generate all-zero vectors in the first iteration, which may cause the gradients of network parameters to become zero. Thus, the network retains its original parameters and keeps generating all-zero vectors, hindering the learning process. To avoid this problem, we utilize the following weight and bias initialization
\begin{align*}
W_{{\bf I}, w}\sim \mathcal{N} (\boldsymbol{\mu}_{\bf I},\boldsymbol{\Sigma}_{\bf I}),\hspace{5pt} W_{{\bf I}, b}=\mathbf{0}, \tag{27}
\end{align*}
where $W_{{\bf I}, w}$ and $W_{{\bf I}, b}$ denote the weight and bias of the linear layer, respectively. $\boldsymbol{\mu}_{\bf I}$ is a vector, where all elements are $\mu_{\bf I}$, and $\boldsymbol{\Sigma}_{\bf I}=\sigma_{\bf I}^2{\bf J}$, where $\bf J$ is the identity matrix. Besides, we set $\mu_{\bf I}>\sigma_{\bf I}^2>0$. This initialization ensures the outputs of these four networks are positive in the first iteration, thus facilitating the learning process. The effectiveness of this initialization is also validated in subsequent experiments.

\section{Meta-learning empowered GNN for adaptive power control}
\label{metagnn algorithm}
% \begin{align*} 
% &\max _{\lbrace{\bf {p}}_{\tau},{\bf {r}}_{\tau},{\bf {s}}_{\tau}\rbrace^{N_T}_{\tau=1}} \hspace{5pt} \sum_{\tau=1}^{N_T} \Big\lbrace {\bf {r}}_{\tau}^T\mathbf{1}_{K_{\tau}} - \frac{\alpha}{2}\Vert {\bf {s}}_{\tau}\Vert _{2}^{2}\Big\rbrace, \tag{5a}\\ 
% &\;\;\;\;\;\;\text{s.t.} \hspace{23pt} {\bf {r}}_{\tau}\leq \mathbb{E} \left[ {\bf {f}}({\bf {H}}, {\bf {p}}_{\tau}({\bf {H}})) \right],\tag{5b} \\ 
% &\;\;\;\; \hspace{40pt} {\bf {r}}_{\tau}\geq {\bf {f}}_{\min } - {\bf {s}}_{\tau}, \tag{5c}\\ 
% &\;\;\;\; \hspace{40pt} {\bf {p}}_{\tau}({\bf {H}}) \in [0,P_{\max }]^{K_{\tau}}, {\bf {s}}_{\tau}\geq {\bf 0}, \forall \tau.\tag{5d}
% \end{align*}
% where $K_{\tau}$ is the number of communication links of the configuration $\tau$. It is worth noting that our constraints (5b)-(5d) are for each configuration, in contrast to those in \cite{zhao2020primal} and \cite{zhao2020fair}, which makes the problem more challenging to solve.
Recall the system model discussed in Sec. \ref{system model}, where the optimization problems are solved only on a single configuration. However, in practice, the learned schemes are required to generalize to novel configurations. This is challenging as acquiring the parameters for a specific configuration $\tau$ through the method discussed in Sec. \ref{gnn algorithm} typically involves numerous iterations and a substantial quantity of channel data samples. Besides, when the configuration changes, the originally trained parameters may suffer severe performance degradation. Additionally, substantial channel data samples are often unavailable in practice. Therefore, in this paper, we propose a meta-learning empowered algorithm to improve the generalization capability of the aforementioned GNN-based primal-dual learning method, which is called \textit{MetaGNN} method. Meta-learning optimizes meta-parameters by training across a variety of tasks, which are then served as the initialization, enhancing the GNNs' ability to rapidly adapt to new configuration and reducing the required number of data samples when applied in real-world scenarios. It is worthy of emphasizing again that the inherent generalization capability of GNNs mitigates the issue of dimensional changes in network parameters when there are variations in the number of communication links, thus enabling the feasibility of meta-learning. As shown in Fig. \ref{metapro}, our algorithm consists of meta-training and adaptation stages, which are discussed in details below. Here, obtaining parameters for aforementioned five GNNs in the wireless network of configuration $\tau$ is considered as the $\tau$th task. Instantaneous channel matrices $\lbrace{\bf H}_t\rbrace_{t=1}^{T_{\tau}}$ drawn from $\mathcal{U}_{\tau}({\bf H})$ are provided as data samples. The training dataset consists of the configuration information and the channel data samples, i.e., $\mathcal{D}^{\text{tr}}_{\tau}=[{\bf N}_{\tau}, \lbrace{\bf H}_t\rbrace_{t=1}^{T_{\tau}}]$. 
\subsubsection{Meta-training Stage}
The meta-training stage aims to obtain meta-parameters via training across a variety of tasks. Considering different numbers of communication links and positions of the transmitters and receivers, a number of configurations with sufficient channel data samples can be generated, denoted as $\mathcal{T}=\{1, \dots, N_T\}$. By calculating the Lagrangian dual functions of each task and combining them, the objective function of meta-learning (\ref{metaobjective}) is obtained. We use $\boldsymbol{\theta}=(\boldsymbol{\theta}^{{\bf {p}}}, \boldsymbol{\theta}^{\bf {r}}, \boldsymbol{\theta}^{\bf {s}}, \boldsymbol{\theta}^{\boldsymbol{\lambda}}, \boldsymbol{\theta}^{\boldsymbol{\mu}})$ to represent meta-parameters. Meta-parameters are firstly updated on each task using the parameterized primal-dual learning method, which is called the inner loop in the field of meta-learning, and then (\ref{metaobjective}) is obtained by summing the Lagrangian dual functions calculated with these updated parameters. This setting differs from the multi-task learning approach, facilitating the meta-parameters' ability to rapidly adapt to new tasks, which is the core principle of MAML \cite{finn2017model}. If we define the updated parameters as 
\begin{align*}
\boldsymbol{\theta}^{A}_{\tau}&:=\boldsymbol{\theta}^{A}+\eta_{A} \nabla _{\boldsymbol{\theta }^{A}}\mathcal{L}_{\tau}(\boldsymbol{\theta}),\hspace{5pt} A\in\{{\bf p}, {\bf r}, {\bf s}\}, \tag{29}
\end{align*}
and
\begin{align*}
\boldsymbol{\theta}^{B}_{\tau}&:=\boldsymbol{\theta}^{B}-\eta_{B}\nabla_{\boldsymbol{\theta}^B}\mathcal{L}_{\tau}(\boldsymbol{\theta}), \hspace{5pt} B\in\{\boldsymbol{\lambda}, \boldsymbol{\mu}\}, \tag{30}
\end{align*}
then the meta-learning problem shown in (\ref{metaobjective}) can be solved by gradient ascent or descent. Notably, these parameters are different from the task-specific parameters mentioned in Sec. \ref{III-A}. Similar to \cite{finn2017model}, for simplicity, we only consider one-step update for analysis, while updating parameters for multiple steps in the inner loop is also feasible. $\boldsymbol{\theta}_{\tau}=(\boldsymbol{\theta}^{\bf p}_{\tau},\boldsymbol{\theta}^{\bf r}_{\tau},\boldsymbol{\theta}^{\bf s}_{\tau},\boldsymbol{\theta}^{\boldsymbol{\lambda}}_{\tau},\boldsymbol{\theta}^{\boldsymbol{\mu}}_{\tau})$ is used to denote the updated parameters of the $\tau$th task. The meta-parameters are updated as follows
\begin{align*}
\boldsymbol{\theta}^A &\leftarrow \boldsymbol{\theta}^A + \varepsilon_A\nabla_{\boldsymbol{\theta}^A}\sum_{\tau=1}^{N_T} \mathcal{L}_{\tau}(\boldsymbol{\theta}_{\tau})\\
&\approx \boldsymbol{\theta}^A+\varepsilon_A\sum_{\tau=1}^{N_T} \nabla_{\boldsymbol{\theta}^{A}_{\tau}}\mathcal{L}_{\tau}(\boldsymbol{\theta}_{\tau})\cdot\left(\mathbb{I}+\eta_{A}\nabla_{\boldsymbol{\theta}^A}^2\mathcal{L}_{\tau}(\boldsymbol{\theta})\right), \tag{31}
\end{align*}
where $A\in \{{\bf p}, {\bf r}, {\bf s}\}$ and similarly, 
\begin{align*}
\boldsymbol{\theta}^{B} &\approx \boldsymbol{\theta}^{B}-\varepsilon_{B}\sum_{\tau=1}^{N_T}\nabla_{\boldsymbol{\theta}^{B}_{\tau}}\mathcal{L}_{\tau}(\boldsymbol{\theta}_{\tau})\cdot\left(\mathbb{I}-\eta_{B}\nabla_{\boldsymbol{\theta}^B}^2\mathcal{L}_{\tau}(\boldsymbol{\theta})\right), \tag{32}
\end{align*}
where $B\in\{\boldsymbol{\lambda}, \boldsymbol{\mu}\}$. $\varepsilon_{A}, A\in\{{\bf p}, {\bf r}, {\bf s}, \boldsymbol{\lambda}, \boldsymbol{\mu}\}$ denotes the step size of meta-learning. The second order derivative terms, i.e., $\nabla_{\boldsymbol{\theta}^{A}}^2\mathcal{L}_{\tau}(\boldsymbol{\theta})$ and $\nabla_{\boldsymbol{\theta}^{B}}^2\mathcal{L}_{\tau}(\boldsymbol{\theta})$, are usually omitted to obtain the first-order MAML (FOMAML) method. It indicates that meta-learning mainly relies on the first-order gradient information to improve the generalization. In this paper, we utilize a variant of FOMAML called Reptile \cite{nichol2018first}, which also incorporates such first-order information while eliminating the requirement of backward process. Notice that Reptile requires multiple steps in the inner loops to obtain the first-order gradient information. The meta-parameters are then updated as
\begin{align*}
\boldsymbol{\theta}^{\bf I}&\leftarrow \boldsymbol{\theta}^{\bf I}+\frac{\varepsilon_{\bf I}}{N_T}\sum_{\tau=1}^{N_T}(\boldsymbol{\theta}^{\bf I}_{\tau}-\boldsymbol{\theta}^{\bf I})/ \eta_{\bf I}, \tag{33}\label{reptileupdate}
\end{align*}
where ${\bf I}\in\{{\bf p}, {\bf r}, {\bf s}, \boldsymbol{\lambda}, \boldsymbol{\mu}\}$. This process is called the outer loop, which combines the results of inner loops to refine the meta-parameters.  
\subsubsection{Adaptation Stage}
\begin{algorithm}[htbp]
\caption{Meta-learning Empowered GNN-based Radio Resource Management (\textit{MetaGNN})}
\label{alg1}
\begin{algorithmic}[1]
\Require
Inner loop learning rates: $(\eta_p,\eta_r,\eta_s,\eta_{\lambda},\eta_{\mu})$, \newline
outer loop learning rates: $(\varepsilon_p,\varepsilon_r,\varepsilon_s,\varepsilon_{\lambda},\varepsilon_{\mu})$, \newline
pre-generated meta-training set $\mathcal{T}$ and the unseen task $\tau_{\text{test}}$. \newline
\rule[1pt]{0.46\textwidth}{0.05em}
\Statex \textbf{Meta-training Stage:}
\State Initialize the meta-parameters $\boldsymbol{\theta}$
% $\boldsymbol{\theta}_0^{\bf p}, \boldsymbol{\theta}_0^{\bf r}, \boldsymbol{\theta}_0^{\bf s}, \boldsymbol{\theta}_0^{\boldsymbol{\lambda}}, \boldsymbol{\theta}_0^{\boldsymbol{\mu}},$ 
randomly. 
\For{outer loop $o\in[1, N_O]$}
\State Sample $N_b$ tasks from the meta-training set as a batch.
\For{task $\tau\in[1, N_b]$}
\State Initialize the GNNs parameters of the $\tau$th task with meta-parameters, i.e.,  $\boldsymbol{\theta}_{\tau, 1}\leftarrow\boldsymbol{\theta}$.
\For{inner loop $i\in[1, N_I]$}
\State Calculate Lagrangian dual function $\mathcal{L}_{\tau}(\boldsymbol{\theta}_{\tau, i})$.
\State Update $\boldsymbol{\theta}_{\tau, i}^{\bf I}$, ${\bf I}\in\{{\bf p}, {\bf r}, {\bf s}, \boldsymbol{\lambda}, \boldsymbol{\mu}\}$ with (\ref{updatep})-(\ref{updatem}).
\EndFor
\State Extract $\boldsymbol{\theta}_{\tau, N_I}$ as updated parameters $\boldsymbol{\theta}_{\tau}$.
\EndFor
\State Update meta-parameters with (\ref{reptileupdate}).
\EndFor\newline
\rule[1pt]{0.46\textwidth}{0.05em}
\Statex \textbf{Adaptation Stage:}
\State Initialize the GNNs with meta-parameters obtained from meta-training stage $\boldsymbol{\theta}_{\text{test}, 1}\leftarrow\boldsymbol{\theta}$.
\For {adapt iteration $i\in[1, N_a]$}
\State Calculate Lagrangian dual function $\mathcal{L}_{\text{test}}(\boldsymbol{\theta}_{\text{test}, i})$.
\State Fine-tune $\boldsymbol{\theta}_{\text{test}, i}^{\bf I}$, ${\bf I}\in\{{\bf p}, {\bf r}, {\bf s}, \boldsymbol{\lambda}, \boldsymbol{\mu}\}$, using (\ref{updatep})-(\ref{updatem}).
\EndFor
\State Extract $\boldsymbol{\theta}_{\text{test}, N_a}$ as the task-specific parameters $\boldsymbol{\theta}_{\text{test}}$.
\end{algorithmic}
\end{algorithm}
The meta-parameters derived from the meta-training stage serve as the initialization of the GNNs for an unseen task $\tau_{\text{test}}$. As shown in Fig. \ref{metapro}, a dataset $\mathcal{D}_{\text{test}}$ that contains configuration information ${\bf N}_{\text{test}}$ and a limited number of channel samples $\lbrace{\bf H}_t\rbrace_{t=1}^{T_{\text{test}}}$ is provided. From the perspective of meta-learning, meta-parameters enable the GNNs to extract crucial information from limited data and fine-tune their parameters for adaptation when confronting unseen tasks. The adaptation stage is similar to the inner loop, in which we fine-tune the GNN parameters using (\ref{updatep})-(\ref{updatem}) within the given dataset. Since the meta-parameters have already encapsulated some structural information about dealing with these primal-dual learning problems, only a few epochs suffice to obtain an efficient resource management scheme. The whole \textit{MetaGNN} algorithm is summarized in Algorithm \ref{alg1}. Here, an extra subscript $i$ is used to indicate the number of steps in the inner loop.

\section{Numerical Result}\label{numerical result}
\begin{figure}[ht!]
\centering
\subfigure[Average mean rate]{
    \includegraphics[width=0.41\textwidth]{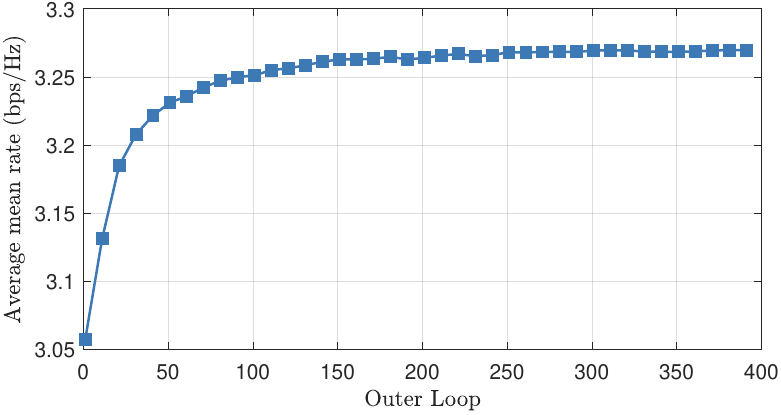} 
}
\subfigure[Average minimum rate]{
    \includegraphics[width=0.41\textwidth]{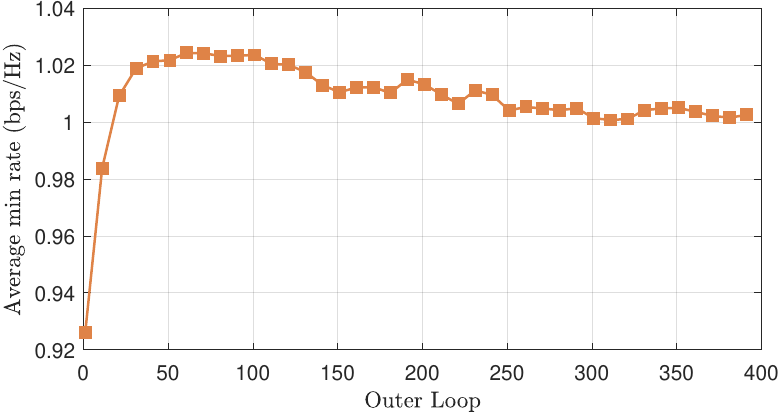} 
}
\caption{The average mean and minimum ergodic rates during meta training stage.}
\label{metatrain}
\end{figure}
In this section, we provide the numerical results of using the proposed MetaGNN method to solve the adaptive power control problems in wireless networks. 
Similar with \cite{navid2022state}, \cite{ivana2023modular} and \cite{eisen2019optimal}, we consider a scenario with $K$ transmitters randomly dropped within a 500 m $\times$ 500 m square area. The receivers are uniformly and randomly dropped around transmitters, ensuring that minimum and maximum pairwise distances of 10 m and 50 m for each transmitter-receiver pair, respectively. The large-scale fading component consists of a log-normal shadowing components with standard deviation $\xi=7$ dB and a dual-slope path loss \cite{navid2023resilient}\cite{zhang2015downlink}. 
\begin{table}[htbp]
\centering
\caption{Major Parameters of meta-training and adaptation}
\resizebox{\linewidth}{!}{
\begin{tabular}{c|c}\hline
Parameter & Value\\\hline
Channel samples for training per task & 2048\\
Channel samples for testing per task & 256\\
Averaging window length $\beta$ & 0.05\\
Slack norm regularization parameter $\alpha$ & 4.0\\
The inner loop learning rate of power net $\eta_p$ & 2e-2\\
The inner loop learning rates of slack and dual nets $\eta_s,\eta_{\mu}$ & 3e-3\\
The outer loop learning rate of power net $\varepsilon_p$ & 1e-2\\
The outer loop learning rates of slack and dual nets $\varepsilon_s,\varepsilon_{\mu}$ & 1e-3\\
The adaptation rate of power net $\eta_p^{a}$ & 4e-3\\
The adaptation rates of slack and dual nets $\eta_s^{a},\eta_{\mu}^{a}$ & 8e-4\\
\hline
\end{tabular}}
\label{tab1}
\end{table}
Rayleigh fading is utilized as small-scale fading. We can then generate multiple instantaneous channel matrices as training samples. The bandwidth of whole network is $W=10$ MHz, the maximum power for transmission is $P_{\max}=10$ dBm and the noise power spectral density is -174 dBm/Hz. We use the LEConv layer in PyTorch Geometric library \cite{fey2019fast}, which aggregates the information from adjacent nodes. Each GNN consists of two hidden layers, both containing 128 neurons and the LeakyReLU is used as the activation function.

\begin{figure}[ht!]
\centering
\subfigure[Mean rate]{
    \includegraphics[width=0.41\textwidth]{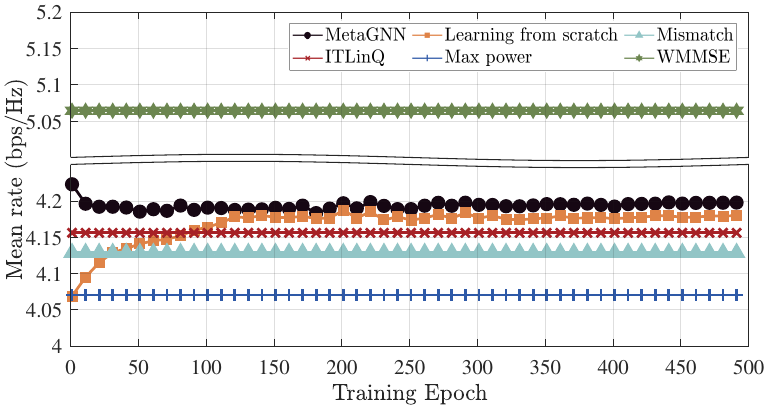} 
}
\subfigure[Minimum rate]{
    \includegraphics[width=0.41\textwidth]{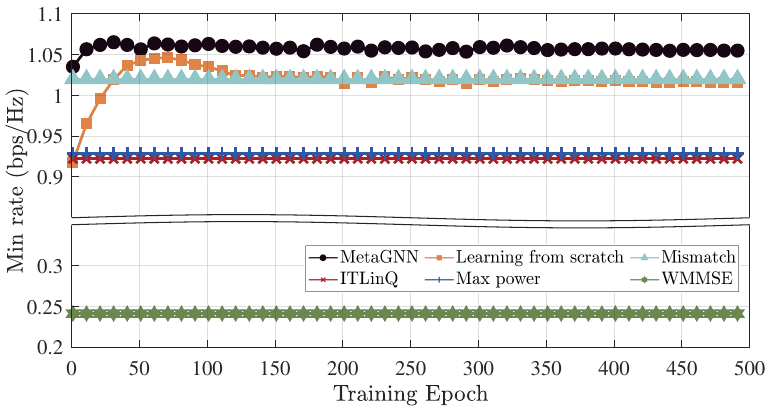} 
}
% Number of users: 8, index: 196.
\caption{The performance comparison during adaptation in a new network configuration $\tau_{\text{test}}$.}
\label{singleconfig}
\end{figure}
In this paper, we mainly consider changing the number of communication links in the network and their positions. Once these two factors are determined, they form a specific configuration. We define the task as obtaining an efficient scheme for adaptive power control in a specific configuration. The major parameters of meta-training and adaptation are shown in Table \ref{tab1}. Besides, based on experiments, the value of ${\bf f}_{\min}$ affects the final performance, but a reasonable and effective setting has yet to be found in this paper. We use a heuristic approach as
\begin{equation*} 
f_{k,\min} = 
\begin{cases} 
\displaystyle \text{round}(\max(\tilde{\bf r}(0))), & \text {if $\max(\tilde{\bf r}(0)) \leq \varsigma$},\\ 
\varsigma, & \text {otherwise},
\end{cases}
\tag{34}
\end{equation*}
where $\varsigma$ is the threshold of the minimum rate that can be selected from training several tasks and round($\cdot$) denotes the numerical operation that implements the conventional rounding rule. According to the initial long-term average rate $\tilde{\bf r}(0)$, we have a general understanding of the channel conditions, with which we can set the $f_{k,\min}$ properly to ensure that the duality gap shown in (\ref{duality gap}) is within an acceptable range. During meta-training, a fixed number of tasks are randomly sampled from the meta-training set per outer loop, with the number as $N_b=32$. The meta-parameters are derived through $N_O=400$ outer loops. For each inner loop, $N_I=20$ epochs are conducted to obtain updated parameters. Following the update of the meta-parameters, we sample 30 testing tasks to evaluate performance. It is noteworthy that we test the GNNs initialized with the updated meta-parameters and roll out 20 epochs in the corresponding testing task for adaptation. The average mean and minimum ergodic rates during meta-training stage are presented in Fig. \ref{metatrain}. From the figure, in the first 100 epochs, the average minimum-rate rapidly increases, surpassing the relaxed minimum rate constraint. Then during 100$\sim$250 epochs, the average mean-rate further improves at the cost of sacrificing some average minimum-rate performance. Lastly, both metrics approximately converge after about 250 epochs.

In order to see how the meta parameter affects the adaptation in a new configuration, we show the mean and minimum ergodic rates of our MetaGNN approaches against several baselines in Fig. \ref{singleconfig}. These baselines are introduced as follows.
\begin{itemize}
    \item \textbf{Learning from scratch}: Use the parameterized primal-dual learning method to train the randomly initialized GNNs, i.e., $\boldsymbol{\theta}^{\bf p},\boldsymbol{\theta}^{\bf s},\boldsymbol{\theta}^{\boldsymbol{\mu}}$, in the considered network configuration for 500 epochs.
    \item \textbf{Mismatch}: Use the parameterized primal-dual learning method to train the random-initialized GNNs in a network configuration with different number of communication links for 500 epochs.
    \item \textbf{ITLinQ} \cite{navid2014itlinq}: This method selects a group of communication links at each time step according to their PF ratios, signal-to-noise ratio (SNR) and interference-to-noise ratio (INR) values, which then transmit with $P_{\max}$. This method takes the minimum rate performance into consideration, hence improving fairness across all links.
    \item \textbf{Max power}: This method lets all communication links transmit with $P_{\max}$ at each time step.
    \item \textbf{WMMSE} \cite{shi2011wmmse}: WMMSE is a popular algorithm for optimizing sum rate in multi-user wireless networks. The iterative procedure is repeated for 100 times here.
\end{itemize}

From Fig. \ref{singleconfig}, our method converges after 20-30 epochs, which achieves adaptation to new network configurations by trading a slight decrease in the mean ergodic rate for an improvement in the minimum ergodic rate, surpassing all baselines except WMMSE. The learning from scratch method starts at a low level and converges after about 200 epochs, which takes nearly 10 times more adaptation time than MetaGNN. During training, the minimum-rate initially rises rapidly and then gradually decreases to further improve the mean-rate. Compared to MetaGNN, the post-convergence performance of both metrics shows varying degrees of decline. From our perspective, for the non-convex power control problem corresponding to this new configuration, the learning from scratch method is more prone to getting stuck in a local optimum, whereas MetaGNN effectively avoids this situation by providing a good initialization. Besides, although the mismatched scheme achieves similar minimum-rate performance to that of learning from scratch, its mean-rate performance obviously declines compared to both the learning from scratch and MetaGNN methods. This indicates that, despite the generalization of GNNs, performance degradation occurs when network changes, demonstrating the importance of meta-learning. Additionally, ITLinQ and Max power methods consider the minimum-rate performance using relatively simple approaches. Their consideration results in severe degradation of mean-rate performance as they are not specifically designed for the considered problem. Particularly, MetaGNN improves mean-rate by 1.2$\%$ and minimum-rate by over 13$\%$ compared to ITLinQ. Lastly, since the WMMSE algorithm aims to maximize the weighted sum rate of the whole network, it performs well in terms of mean-rate. In contrast, its minimum-rate performance is very poor because it sacrifices the performance of heavily interfered users to enhance the sum rate, which is particularly unfair.

\begin{figure}[ht!]
\centering
\subfigure[Mean rate]{
\includegraphics[width=0.44\textwidth]{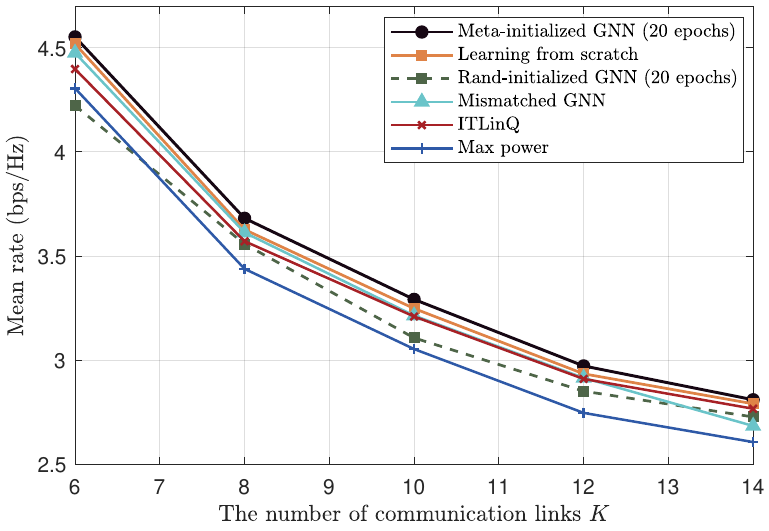} 
}
\subfigure[Minimum rate]{
\includegraphics[width=0.44\textwidth]{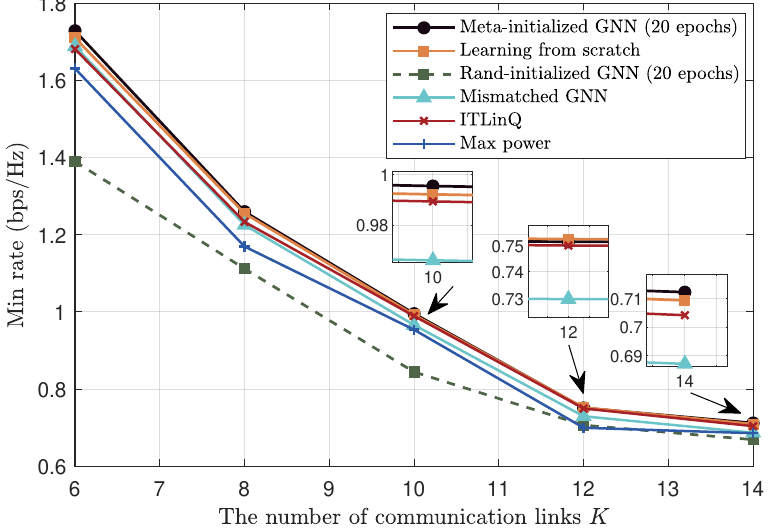} 
}
\caption{The performance comparison of MetaGNN method against several baselines with varying numbers of communication links $K$.}
\label{metacomp}
\end{figure}
\begin{figure}[ht!]
\centering
\subfigure[Mean rate]{
    \includegraphics[width=0.42\textwidth]{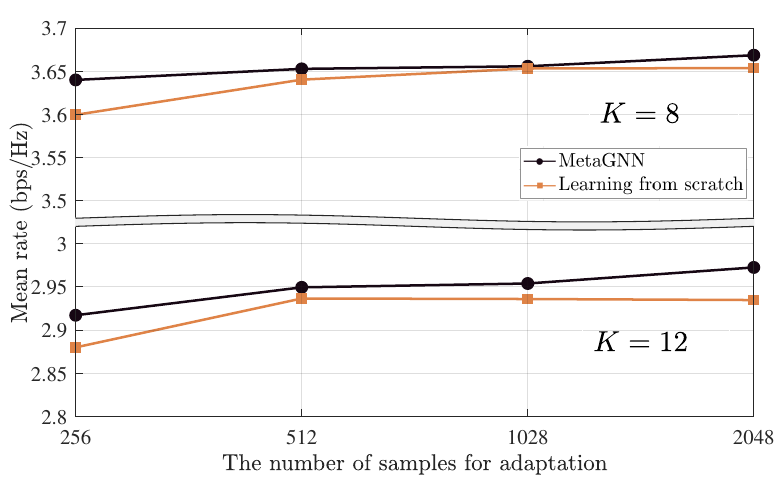} 
}
\subfigure[Minimum rate]{
    \includegraphics[width=0.42\textwidth]{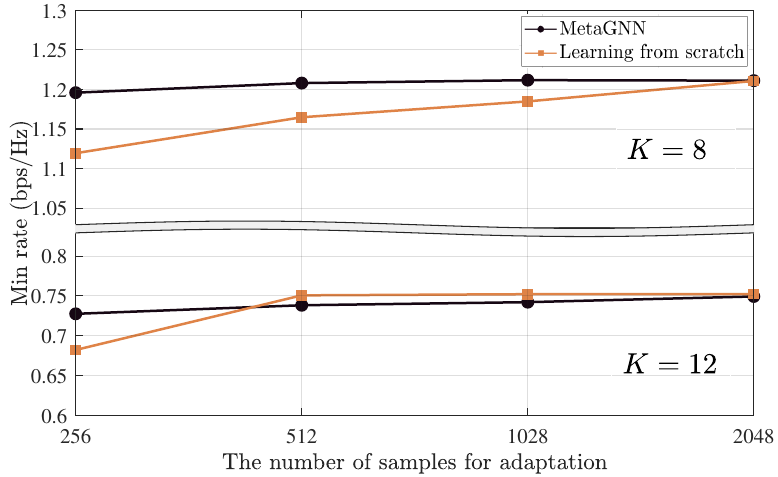} 
}
\caption{The performance comparison of MetaGNN and learning from scratch method with varying number of channel data samples in adaptation.}
\label{metasample}
\end{figure}
Although MetaGNN outperforms almost all baselines in both metrics for a specific configuration considered in Fig. \ref{singleconfig}, the experiments reveal that this is not consistently the case across all configurations. Therefore, Fig. \ref{metacomp} illustrates the average performance of our method compared to several baselines across multiple configurations with varying numbers of users, providing a more comprehensive evaluation of its effectiveness. Our method is labeled ``Meta-initialized GNN (20 epochs)'' here, which means that the GNNs are initialized with meta-parameters and roll out 20 epochs in the new configurations. An extra baseline, which is labeled ``Rand-initialized GNN (20 epochs)'', is also considered here. This baseline initializes the parameters of GNNs randomly and adapts them in the new configurations for 20 epochs, which is essentially a learning from scratch method, but with fewer training epochs. From Fig. \ref{metacomp}, MetaGNN and learning from scratch methods exhibit comparable performance across both metrics, with MetaGNN showing a slight advantage. This improvement can likely be attributed to the benefits of meta-parameters, which help in avoiding local optimum during training. Mismatched GNN scheme suffers performance degradation in both metrics, indicating the necessity of using meta-learning or other methods to improve its generalization to deal with the variation of network configurations. Besides, when compared to the ITLinQ method, MetaGNN demonstrates similar minimum-rate performance. Since MetaGNN only requires the minimum-rate to exceed the relaxed constraint, i.e., ${\bf f}_{\min}-{\bf s}$, it allows for a trade-off by slightly sacrificing minimum-rate performance to further enhance the sum rate once the constraint is satisfied. Consequently, MetaGNN outperforms ITLinQ by about 3.48$\%$ in the mean-rate performance, demonstrating that we achieve a superior trade-off compared to ITLinQ. Moreover, the rand-initialized method that roll out for 20 epochs has the worst minimum-rate performance among these baselines while the mean-rate performance can hardly surpass the mismatched scheme. Therefore, the superior performance of the meta-initialized GNN method is mainly derived from meta-parameters, which are obtained from training across a variety of network configurations in the meta-training stage.

Apart from the ability of MetaGNN to speed up adaptation to new network configurations, it can also reduce the amount of channel data samples required for effective adaptation. To validate this advantage, we conduct experiments comparing the performance of MetaGNN and learning from scratch methods in multiple new network configurations, evaluating their average performance with the varying number of available channel data samples, the results are shown in Fig. \ref{metasample}. Although we only show the performance in wireless networks with $K=8$ and $K=12$ communication links, similar results can be observed in networks with different numbers of communication links. Here, MetaGNN refers to the GNNs initialized with meta-parameters and adapted for 20 epochs in new configurations. 
\begin{figure}[ht!]
\centering
\subfigure[Mean rate]{
\includegraphics[width=0.42\textwidth]{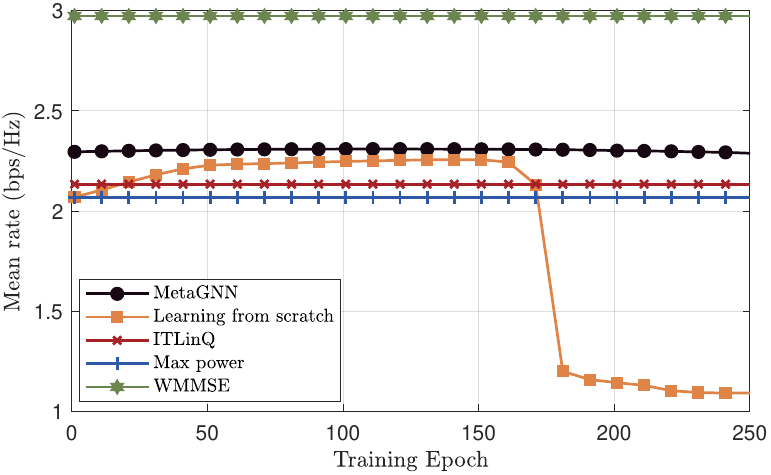} 
}
\subfigure[Minimum rate]{
\includegraphics[width=0.42\textwidth]{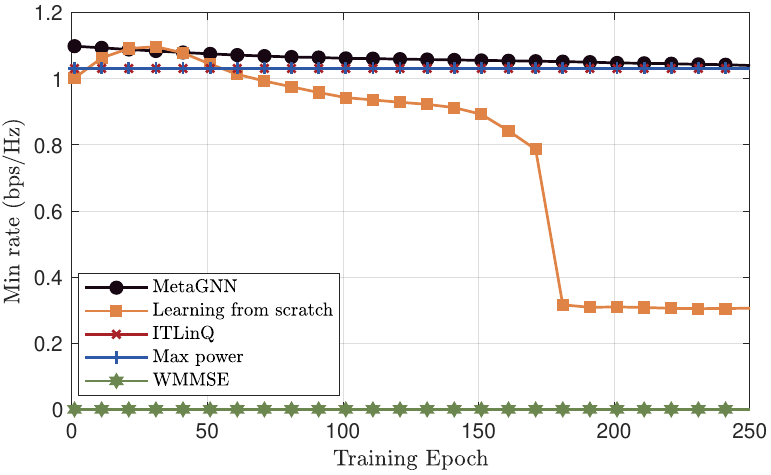} 
}
\caption{The performance comparison of MetaGNN and baselines during adaptation in a new network configuration with only 256 channel data samples.}
\label{metafewsample}
\end{figure}
As illustrated in Fig. \ref{metasample}, the MetaGNN method maintains nearly constant minimum-rate performance as the number of channel data samples decreases, with only a slight reduction in mean-rate performance. In contrast, learning from scratch method only achieves desirable performance with sufficient data samples. When the number of provided data samples is reduced, particularly to 256 data samples, it will experience a significant decline in both performance metrics. From our perspective, updating the GNNs with these limited data samples for hundreds of iterations may introduce significant estimation bias regarding the true channel distribution and trap them into overfitting, further hindering the performance and generalization of GNNs. To further demonstrate the efficiency of MetaGNN in adapting to network configurations with limited data samples, we choose a specific configuration to show the learning processes of these two methods in Fig. \ref{metafewsample}. From Fig. \ref{metafewsample}, we find that during the first 150 epochs, the mean-rate of learning from scratch increases slowly, while the minimum-rate initially rises but then gradually declines, until being worse than that of the Max power method. This decline is probably due to estimation bias issue caused by the limited data samples. After 150 epochs, the method appears to suffer from overfitting, leading to significant performance degradation. MetaGNN, on the contrary, has minimal performance degradation within the first 250 epochs, except for a slight decrease in the minimum-rate. Besides, since MetaGNN typically requires only dozens of iterations in new configuration, overfitting can be naturally avoided. Conversely, learning from scratch fails to achieve desirable performance throughout the whole adaptation process, indicating its ineffectiveness in these configurations with limited data samples, which explains the performance degradation shown in Fig. \ref{metasample}. In summary, MetaGNN can not only speed up the adaptation process but also reduce the required training samples, showing its great potential in RRM problems.

% \textcolor{red}{The change of the training and adaptation stage. 20 epochs not only reduce the cost of the adaptation stage, and also avoid the risk of overfitting. Besides, it can help demonstrate that our method can not only speed up the convergence and also reduce the required training samples.}
% \textcolor{red}{To demonstrate the reason why learning from scratch within fewer training samples has a poor performance. There are some configuration where learning from scratch cannot learn well, our metaGNN may also fall down to lower level, but with a longer epoch. Learning from scratch cannot achieve the best results.}

% \section*{Acknowledgment}
%Dr. Reveryrand would like to acknowledge the funding by XLIM, Limoges, France. 
% The authors would like to thank Dr. David Root and Dr. Jean-Pierre Teyssier at Agilent Technologies for the loan of the time-domain nonlinear measurement equipment and TriQuint Semiconductor for the donation of the transistors. 
\section{Conclusion}
\label{conclusion}
In dynamic wireless networks, meta-learning enables rapid adaptation of adaptive power control schemes and reduces the number of required data samples. In this paper, we first formulate a resilient optimization problem to maximize sum ergodic rates with per-user minimum-rate constraints and derive its Lagrangian dual problem, which is solved by a parameterized primal-dual learning method. As this scheme is required to generalize to different network configurations, we investigate the integration of meta-learning and parameterized primal-dual learning to further enhance the generalization and scalability. However, existing methods suffer from the varying dimensionality due to the changes of network configurations, such as the number of links. To address this problem, we choose the GNNs as the parametrization because of their great scalability, making meta-learning feasible. Afterwards, we propose a meta-learning empowered GNN method for RRM and simulation results demonstrate the effectiveness and scalability of our algorithm. Additionally, the reduction of required CSI information and the integration of more advanced meta-learning methods to enable the networks to adapt to new configurations without fine-tuning will be further explored in future works.

\appendices
\section{Proof of Theorem 1}
\label{proof}
% \textcolor{red}{[2] consider move this prove to the meta-learning objective, it naturally introduce the expectation over configurations.}

Here, we omit the subscript $\tau$ to keep the proof uncluttered. For convenience, we define the unparameterized power policy, primal variables, and dual variables as ${\bf p}$, ${\tilde{\bf {x}}}=[{\bf r}, {\bf s}] $, and $\boldsymbol{\gamma}=[\boldsymbol{\lambda}, \boldsymbol{\mu}]$. Further define $\boldsymbol{\theta }^{\bf x}=[\boldsymbol{\theta }^{\bf r}, \boldsymbol{\theta }^{\bf s}]$ and $\boldsymbol{\theta }^{\boldsymbol{\gamma}}=[\boldsymbol{\theta }^{\boldsymbol{\lambda}}, \boldsymbol{\theta }^{\boldsymbol{\mu}}]$ to collect the parameterized primal parameters and dual parameters. Then we can rewrite (\ref{p-lag}) as
\begin{align*} 
{\mathcal {L}}_\theta (\boldsymbol{\theta }^{\bf p}, \boldsymbol{\theta }^{\bf x}, \boldsymbol{\theta }^{\boldsymbol{\gamma}}) &= {\mathcal {F}}(\boldsymbol{\theta }^{\bf x}) - \boldsymbol{\Gamma}(\boldsymbol{\theta}^{\boldsymbol{\gamma}})^{T} {\mathcal {G}}(\boldsymbol{\theta }^{\bf x}; \boldsymbol{\theta }^{\bf p}), 
\tag{35}
\end{align*}
where ${\mathcal {F}}(\boldsymbol{\theta }^{\bf x})={\bf {r}}({\bf{N}}; \boldsymbol{\theta}^{\bf r})^T\mathbf{1}_K - \frac{\alpha }{2}\Vert {\bf {s}}({\bf{N}}; \boldsymbol{\theta}^{\bf s})\Vert _{2}^{2}$, $\boldsymbol{\Gamma}(\boldsymbol{\theta}^{\boldsymbol{\gamma}})=[\boldsymbol{\lambda}({\bf{N}};\boldsymbol{\theta}^{\boldsymbol{\lambda}});\boldsymbol{\mu}({\bf{N}};\boldsymbol{\theta}^{\boldsymbol{\mu}})]$, and ${\mathcal {G}}(\boldsymbol{\theta }^{\bf x}; \boldsymbol{\theta }^{\bf p})=[{\bf {r}}({\bf{N}}; \boldsymbol{\theta}^{\bf {r}})- \mathbb{E}_{{\bf {H}}}[ {\bf {f}}({\bf {H}}, {\bf {p}}({\bf {H}};\boldsymbol{\theta }^{{\bf {p}}}))];{\bf {f}}_{\min } - {\bf {s}}({\bf{N}}; \boldsymbol{\theta}^{\bf {s}})- {\bf {r}}({\bf{N}}; \boldsymbol{\theta}^{\bf {r}})]$.

The optimal parameterized dual value $D_\theta^*$ is obtained from the solution of problem (\ref{dp_para}). Given the fact that the parameterized primal networks are the subsets of the unparameterized classes of primal variables, we have
\begin{align*} 
D_\theta ^* &= \min _{\boldsymbol{\theta }^{\boldsymbol{\gamma}}} \max _{\boldsymbol{\theta }^{\bf p}, \boldsymbol{\theta }^{\bf x}} \left[ {\mathcal {F}}(\boldsymbol{\theta }^{\bf x}) - \boldsymbol{\Gamma}(\boldsymbol{\theta}^{\boldsymbol{\gamma}})^{T} {\mathcal {G}}(\boldsymbol{\theta }^{\bf x}; \boldsymbol{\theta }^{\bf p}) \right], \tag{36}\label{primalproblem}\\
&\leq \min _{\boldsymbol{\theta }^{\boldsymbol{\gamma}}} \max _{{\bf {p}}, {\tilde{\bf {x}}}} \left[ {\mathcal {F}}({\tilde{\bf {x}} }) - \boldsymbol{\Gamma}(\boldsymbol{\theta}^{\boldsymbol{\gamma}})^{T} {\mathcal {G}}({\tilde{\bf {x}} }; {\bf {p}}) \right], \tag{37}\label{subsetupper}
\end{align*}
where ${\mathcal {F}}({\tilde{\bf {x}} })={\bf {r}}^T\mathbf{1}_K - \frac{\alpha }{2}\Vert {\bf {s}}\Vert _{2}^{2}$ and ${\mathcal {G}}({\tilde{\bf {x}} }; {\bf {p}})=[{\bf {r}}- \mathbb{E}_{{\bf {H}}} [ {\bf {f}}({\bf {H}}, {\bf {p}}({\bf {H}}))]; {\bf {f}}_{\min } - {\bf {s}}- {\bf {r}}]$. We add and subtract $\boldsymbol{\gamma}^T{\mathcal {G}}({\tilde{\bf {x}} }; {\bf {p}})$ to and from the right hand side of (\ref{subsetupper}) and obtain
\begin{align*}
D_\theta ^*&\leq \min _{\boldsymbol{\theta }^{\boldsymbol{\gamma}}} \max _{{\bf {p}}, {\tilde{\bf {x}}}} \left\{ {\mathcal {F}}({\tilde{\bf {x}} }) - \boldsymbol{\gamma}^T{\mathcal {G}}({\tilde{\bf {x}} }; {\bf {p}})+ (\boldsymbol{\gamma}-\boldsymbol{\Gamma}(\boldsymbol{\theta}^{\boldsymbol{\gamma}}))^{T} {\mathcal {G}}({\tilde{\bf {x}} }; {\bf {p}}) \right\},\\
&=\max _{{\bf {p}}, {\tilde{\bf {x}}}}\left\{{\mathcal {F}}({\tilde{\bf {x}} })-\boldsymbol{\gamma}^T{\mathcal {G}}({\tilde{\bf {x}}};{\bf {p}})+\min _{\boldsymbol{\theta }^{\boldsymbol{\gamma}}}(\boldsymbol{\gamma}-\boldsymbol{\Gamma}(\boldsymbol{\theta}^{\boldsymbol{\gamma}}))^{T} {\mathcal {G}}({\tilde{\bf {x}} }; {\bf {p}})\right\}.\tag{38}
\end{align*}
When we obtain the optimal ${\bf p}^*$ and $\tilde{\bf x}^*$, then the value of ${\mathcal {G}}({\tilde{\bf {x}}^* }; {\bf {p}^*})$ should be non-positive. Therefore, for any non-negative dual variables $\boldsymbol{\gamma}$, we can always find the parameters $\boldsymbol{\theta}^{\gamma}$ to make $(\boldsymbol{\gamma}-\boldsymbol{\Gamma}(\boldsymbol{\theta}^{\boldsymbol{\gamma}}))^{T} {\mathcal {G}}({\tilde{\bf {x}} }; {\bf {p}})$ non-positive. Hence, we can obtain
\begin{align*}
D_\theta ^*&\leq\min _{\boldsymbol{\gamma}}\max _{{\bf {p}}, {\tilde{\bf {x}}}}[{\mathcal {F}}({\tilde{\bf {x}} })-\boldsymbol{\gamma}^T{\mathcal {G}}({\tilde{\bf {x}}};{\bf {p}})]=D^*=P^*. \tag{39}
\end{align*}

As for the lower bound on $D_\theta ^*$, we add and subtract $\boldsymbol{\Gamma}(\boldsymbol{\theta}^{\boldsymbol{\gamma}})^{T} {\mathcal {G}}(\boldsymbol{\theta }^{\bf x}; {\bf p})$ to and from the right hand side of (\ref{primalproblem}) at first and obtain
\begin{align*} 
D_\theta ^* &= \min _{\boldsymbol{\theta }^{\boldsymbol{\gamma}}}\max _{\boldsymbol{\theta }^{\bf x}}  \Big\lbrace{\mathcal {F}}(\boldsymbol{\theta }^{\bf x}) - \boldsymbol{\Gamma}(\boldsymbol{\theta}^{\boldsymbol{\gamma}})^{T} {\mathcal {G}}(\boldsymbol{\theta }^{\bf x}; {\bf p})\\
&\quad\quad -\min_{\boldsymbol{\theta }^{\bf p} }\left[\boldsymbol{\Gamma} (\boldsymbol{\theta}^{\boldsymbol{\gamma}})^{T} ({\mathcal {G}}(\boldsymbol{\theta }^{\bf x}; \boldsymbol{\theta }^{\bf p}) - {\mathcal {G}}(\boldsymbol{\theta }^{\bf x}; {\bf p}) )\right]\Big\rbrace, \tag{40}\label{lb_1}
\end{align*}

Define the term $\Delta_{\boldsymbol{\theta }^{\bf p}}=\boldsymbol{\Gamma} (\boldsymbol{\theta}^{\boldsymbol{\gamma}})^{T} ({\mathcal {G}}(\boldsymbol{\theta }^{\bf x}; \boldsymbol{\theta }^{\bf p}) - {\mathcal{G}}(\boldsymbol{\theta }^{\bf x}; {\bf p}))$, then (\ref{lb_1}) can be rewritten as
\begin{align*} 
D_\theta ^* &= \min _{\boldsymbol{\theta }^{\boldsymbol{\gamma}}}\max _{\boldsymbol{\theta }^{\bf x}} \left\{ {\mathcal {F}}(\boldsymbol{\theta }^{\bf x}) - \boldsymbol{\Gamma}(\boldsymbol{\theta}^{\boldsymbol{\gamma}})^{T} {\mathcal {G}}(\boldsymbol{\theta }^{\bf x}; {\bf p})
-\min_{\boldsymbol{\theta }^{\bf p}} \Delta_{\boldsymbol{\theta }^{\bf p}}\right\} \\
&\geq \min _{\boldsymbol{\theta }^{\boldsymbol{\gamma}}}\max _{\boldsymbol{\theta }^{\bf x}} \left\{ {\mathcal {F}}(\boldsymbol{\theta }^{\bf x}) - \boldsymbol{\Gamma}(\boldsymbol{\theta}^{\boldsymbol{\gamma}})^{T} {\mathcal {G}}(\boldsymbol{\theta }^{\bf x}; {\bf p})
-\min_{\boldsymbol{\theta }^{\bf p}} |\Delta_{\boldsymbol{\theta }^{\bf p}}|\right\}.\tag{41}\label{lb_2}
\end{align*}

Then our target is to find an upper bound for $\min_{\boldsymbol{\theta}^{\bf p}}|\Delta_{\boldsymbol{\theta }^{\bf p}}|$. Using H$\ddot{\text{o}}$lder's inequality, we have
\begin{align*} 
\min_{\boldsymbol{\theta }^{\bf p}} |\Delta_{\boldsymbol{\theta }^{\bf p}}| &\leq \min_{\boldsymbol{\theta }^{\bf p}} \left\Vert \boldsymbol{\Gamma}(\boldsymbol{\theta}^{\boldsymbol{\gamma}})\right\Vert _{1} \left\Vert {\mathcal {G}}(\boldsymbol{\theta }^{\bf x}; \boldsymbol{\theta }^{\bf p}) - {\mathcal {G}}(\boldsymbol{\theta }^{\bf x}; {\bf {p}})\right\Vert _{\infty}.\tag{42}\label{lb_3}
\end{align*}
By further applying the Lipschitz continuity stated in Assumption 3, we can draw a upper bound as
\begin{align*}
\min_{\boldsymbol{\theta }^{\bf p}} |\Delta_{\boldsymbol{\theta }^{\bf p}}|&\leq \min_{\boldsymbol{\theta }^{\bf p}}  L_{f}\left\Vert \boldsymbol{\Gamma}(\boldsymbol{\theta}^{\boldsymbol{\gamma}})\right\Vert _{1}\mathbb{E}_{{\bf {H}}}\left\Vert {\bf {p}}({\bf {H}}; \boldsymbol{\theta }^{\bf p}) - {\bf {p}}({\bf {H}})\right\Vert _{\infty}.\tag{43}\label{lb_4}
\end{align*}
Besides, we consider that ${\bf {p}}({\bf {H}}; \boldsymbol{\theta }^{\bf p})$ is a $\epsilon_{p}$-universal parameterization according to Assumption 4. Then we obtain
\begin{align*}
\min_{\boldsymbol{\theta }^{\bf p}} |\Delta_{\boldsymbol{\theta }^{\bf p}}|&\leq \epsilon _{p} L_{f}\left\Vert \boldsymbol{\Gamma}(\boldsymbol{\theta}^{\boldsymbol{\gamma}})\right\Vert _{1}.\tag{44}\label{lb_5}
\end{align*}

Combining (\ref{lb_2}) and (\ref{lb_5}), we have
\begin{align*} 
D_\theta ^* &\geq \min _{\boldsymbol{\theta }^{\boldsymbol{\gamma}}}\max _{\boldsymbol{\theta }^{\bf x}} \left\{ {\mathcal {F}}(\boldsymbol{\theta }^{\bf x}) - \boldsymbol{\Gamma}(\boldsymbol{\theta}^{\boldsymbol{\gamma}})^{T} {\mathcal {G}}(\boldsymbol{\theta }^{\bf x}; {\bf p})
-\epsilon _{p} L_{f}\left\Vert \boldsymbol{\Gamma}(\boldsymbol{\theta}^{\boldsymbol{\gamma}})\right\Vert _{1}\right\}.\tag{45}\label{lb_6}
\end{align*}
As the parameterized dual variables are the subsets of the unparameterized classes of dual variables, we can continue (\ref{lb_6}) as
\begin{align*}
D_\theta ^*&\geq \min _{\boldsymbol{\gamma}}\max _{\boldsymbol{\theta }^{\bf x}} \left[{\mathcal {F}}(\boldsymbol{\theta }^{\bf x})-\boldsymbol{\gamma}^T{\mathcal {G}}(\boldsymbol{\theta }^{\bf x}; {\bf p})-\epsilon _{p} L_{f}\left\Vert \boldsymbol{\gamma}\right\Vert _{1}\right].\tag{46}\label{lb_7}
\end{align*}
Here, we add and subtract ${\mathcal{F}}(\tilde{\bf {x}}) - \boldsymbol{\gamma}^T{\mathcal {G}}(\tilde{\bf {x}}; {\bf p})$ to and from the right hand side of (\ref{lb_7}) and obtain
\begin{align*}
D_\theta ^*&\geq \min _{\boldsymbol{\gamma}}\max_{\boldsymbol{\theta }^{\bf x}} \lbrace{\mathcal{F}}(\boldsymbol{\theta }^{\bf x}) - {\mathcal{F}}(\tilde{\bf {x}}) + {\mathcal{F}}(\tilde{\bf {x}}) - \boldsymbol{\gamma}^T{\mathcal {G}}(\tilde{\bf {x}}; {\bf p})\\
&\quad\quad\quad\quad + \boldsymbol{\gamma}^T{\mathcal {G}}(\tilde{\bf {x}}; {\bf p}) - \boldsymbol{\gamma}^T{\mathcal {G}}(\boldsymbol{\theta }^{\bf x}; {\bf p})-\epsilon _{p} L_{f}\left\Vert \boldsymbol{\gamma}\right\Vert _{1} \rbrace\\
&=\min _{\boldsymbol{\gamma}} \Big\lbrace {\mathcal{F}}(\tilde{\bf {x}}) - \boldsymbol{\gamma}^T{\mathcal {G}}(\tilde{\bf {x}}; {\bf p}) - \epsilon _{p} L_{f}\left\Vert \boldsymbol{\gamma}\right\Vert _{1}-\\
&\quad  \min_{\boldsymbol{\theta }^{\bf x}}\left[{\mathcal{F}}(\tilde{\bf {x}}) - \boldsymbol{\gamma}^T{\mathcal {G}}(\tilde{\bf {x}}; {\bf p})-\left[{\mathcal {F}}(\boldsymbol{\theta }^{\bf x})-\boldsymbol{\gamma}^T{\mathcal {G}}(\boldsymbol{\theta }^{\bf x}; {\bf p})\right]\right]\Big\rbrace.\tag{47}\label{lb_8}
\end{align*}

Define the term $\Delta_{\boldsymbol{\theta }^{\bf x}}$ as
\begin{align*}
\Delta_{\boldsymbol{\theta }^{\bf x}}={\mathcal{F}}(\tilde{\bf {x}}) - \boldsymbol{\gamma}^T{\mathcal {G}}(\tilde{\bf {x}}; {\bf p})-{\mathcal {F}}(\boldsymbol{\theta }^{\bf x})+\boldsymbol{\gamma}^T{\mathcal {G}}(\boldsymbol{\theta }^{\bf x}; {\bf p}), \tag{48}
\end{align*}
and similar to (\ref{lb_2}), we have
\begin{align*}
D_{\theta}^*&\geq \min _{\boldsymbol{\gamma}} \Big\lbrace {\mathcal{F}}(\tilde{\bf {x}}) - \boldsymbol{\gamma}^T{\mathcal {G}}(\tilde{\bf {x}}; {\bf p}) - \epsilon _{p} L_{f}\left\Vert \boldsymbol{\gamma}\right\Vert _{1}-\min_{\boldsymbol{\theta}^{\bf x}}|\Delta_{\boldsymbol{\theta }^{\bf x}}|\Big\rbrace.\tag{49}
\end{align*}
The absolute value of $\Delta_{\boldsymbol{\theta }^{\bf x}}$ can be represented as
\begin{align*}
|\Delta_{\boldsymbol{\theta }^{\bf x}}|&=|({\bf r}-{\bf r}({\bf N};  \boldsymbol{\theta}^{\bf {r}}))^T\mathbf{1}_K-(\frac{\alpha }{2}\Vert {\bf {s}}\Vert _{2}^{2}-\frac{\alpha }{2}\Vert {\bf {s}}({\bf{N}}; \boldsymbol{\theta}^{\bf s})\Vert _{2}^{2})\\
&\quad\quad\quad -\boldsymbol{\gamma}^T\mathcal{G}^{'}(\boldsymbol{\theta}^{\bf x})|,\tag{50}\label{lb_9}
\end{align*}
where $\mathcal{G}^{'}(\boldsymbol{\theta}^{\bf x})=[{\bf r}-{\bf r}({\bf N};  \boldsymbol{\theta}^{\bf {r}});{\bf r}({\bf N};  \boldsymbol{\theta}^{\bf {r}})-{\bf r}+{\bf s}({\bf N};  \boldsymbol{\theta}^{\bf {s}})-{\bf s}]$.

By using the modulus inequality and H$\ddot{\text{o}}$lder's inequality, we draw an upper bound of (\ref{lb_9}), which is written as
\begin{align*}
\min_{\boldsymbol{\theta}^{\bf x}}|\Delta_{\boldsymbol{\theta }^{\bf x}}|&\leq \min_{\boldsymbol{\theta}^{\bf x}}\Big\lbrace|({\bf r}-{\bf r}({\bf N};  \boldsymbol{\theta}^{\bf {r}}))^T\mathbf{1}_K|\\
&+|\frac{\alpha }{2}\Vert {\bf {s}}\Vert _{2}^{2}-\frac{\alpha }{2}\Vert {\bf {s}}({\bf{N}}; \boldsymbol{\theta}^{s})\Vert _{2}^{2}|\\ &+\Vert\boldsymbol{\gamma}\Vert_{1}(\Vert {\bf r}-{\bf r}({\bf N};\boldsymbol{\theta}^{\bf {r}})\Vert_{\infty}+\Vert {\bf s}-{\bf s}({\bf N};\boldsymbol{\theta}^{\bf {s}})\Vert_{\infty})\Big\rbrace.\tag{51}\label{lb_10}
\end{align*}

Obviously, we have $\min_{\boldsymbol{\theta}^{\bf r}}|({\bf r}-{\bf r}({\bf N};  \boldsymbol{\theta}^{\bf {r}}))^T\mathbf{1}_K|\leq \min_{\boldsymbol{\theta}^{\bf r}}K\Vert {\bf r}-{\bf r}({\bf N};\boldsymbol{\theta}^{\bf {r}})\Vert_{\infty}$, where $K$ is the dimension of primal variables $\bf r$ and $\bf s$. We also have 
\begin{align*}
\min_{\boldsymbol{\theta}^{\bf s}}\frac{\alpha }{2}|\Vert {\bf {s}}\Vert _{2}^{2}-\Vert {\bf {s}}({\bf{N}}; \boldsymbol{\theta}^{s})\Vert _{2}^{2}|&\leq\min_{\boldsymbol{\theta}^{\bf s}}\frac{\alpha K}{2}\Vert{\bf s}^2-{\bf s}^2({\bf N};\boldsymbol{\theta}^{\bf {s}})\Vert_{\infty}. \tag{52}
\end{align*}
When we obtain optimal ${\bf s}^*$ and $\boldsymbol{\theta}^{{\bf s}, *}$, we should have $\Vert {\bf s}^*+{\bf s}({\bf N};\boldsymbol{\theta}^{{\bf s},*})\Vert_{\infty}\leq2f_{\min}$ to make problem meaningful. Therefore, we have $\min_{\boldsymbol{\theta}^{\bf s}}\frac{\alpha }{2}|\Vert {\bf {s}}\Vert _{2}^{2}-\Vert {\bf {s}}({\bf{N}}; \boldsymbol{\theta}^{s})\Vert _{2}^{2}|\leq\min_{\boldsymbol{\theta}^{\bf s}}\alpha Kf_{\min}\Vert{\bf s}-{\bf s}({\bf N};\boldsymbol{\theta}^{\bf {s}})\Vert_{\infty}$.
By applying Assumption 5, the upper bound in (\ref{lb_10}) can be turned into
\begin{align*}
\min_{\boldsymbol{\theta}^{\bf x}}|\Delta_{\boldsymbol{\theta }^{\bf x}}|&\leq (K+\Vert\boldsymbol{\gamma}\Vert_{1})\epsilon_r + (\alpha Kf_{\min}+\Vert\boldsymbol{\gamma}\Vert_{1})\epsilon_s. \tag{53}\label{lb_11}
\end{align*}

Combining (\ref{lb_11}) with (\ref{lb_8}), we have 
\begin{align*}
D_\theta ^*&\geq \min _{\boldsymbol{\gamma}} \Big\lbrace{\mathcal{F}}(\tilde{\bf {x}}) - \boldsymbol{\gamma}^T{\mathcal {G}}(\tilde{\bf {x}}; {\bf p}) - \epsilon _{p} L_{f}\left\Vert \boldsymbol{\gamma}\right\Vert _{1}\\
&\quad\quad\quad -(K+\Vert\boldsymbol{\gamma}\Vert_{1})\epsilon_r - (\alpha Kf_{\min}+\Vert\boldsymbol{\gamma}\Vert_{1})\epsilon_s\Big\rbrace. \tag{54}
\end{align*}
Therefore, we have $D_{\theta}^*\geq P^*-\epsilon _{p} L_{f}\left\Vert \boldsymbol{\gamma}^*\right\Vert _{1}-(K+\Vert\boldsymbol{\gamma}^*\Vert_{1})\epsilon_r - (\alpha Kf_{\min}+\Vert\boldsymbol{\gamma}^*\Vert_{1})\epsilon_s$. This completes the proof.

\ifCLASSOPTIONcaptionsoff
  \newpage
\fi

\bibliographystyle{IEEEtran}
\bibliography{IEEEabrv,Bibliography}
%\end{thebibliography}

% ==== SWITCH OFF the BIO for submission
% \begin{IEEEbiography}[{\includegraphics[width=1in,height=1.25in,clip,keepaspectratio]{photo/hk.png}}]{Kai Huang}
% (Graduate Student Member, IEEE) received the B.S. degree in information engineering from Southeast University, Nanjing, China, in 2022. He is currently pursuing the M.S. degree with the School of Information Science and Engineering, Southeast University. His current research interests mainly include wireless communications, resource allocation and reinforcement learning.
% \end{IEEEbiography}

%% insert where needed to balance the two columns on the last page with
%%\newpage

% ==== SWITCH OFF the BIO for submission
% ==== SWITCH OFF the BIO for submission


\vfill
\end{document}